\documentclass[report,jgrga]{agutex}
\usepackage{graphicx}
\usepackage{amsmath}
\usepackage{textcomp}
\usepackage[version=3]{mhchem} 

\authorrunninghead{WORDSWORTH ET AL}

\titlerunninghead{3D MODEL COMPARISONS OF EARLY MARS}

\begin{document}

\title{Comparison of ``warm and wet'' and ``cold and icy'' scenarios for early Mars in a 3D climate model}

\authors{Robin~D.~Wordsworth\altaffilmark{1}}
\altaffiltext{1}{Paulson School of Engineering and Applied Sciences, Harvard University, Cambridge, MA 02138, USA}
\authors{Laura~Kerber\altaffilmark{2}}
\altaffiltext{2}{Jet Propulsion Laboratory, California Institute of Technology, Pasadena, CA 91109, USA}
\authors{Raymond~T.~Pierrehumbert\altaffilmark{3}}
\altaffiltext{3}{Department of the Geophysical Sciences, University of Chicago, Chicago, IL 60637, USA}
\authors{Francois~Forget\altaffilmark{4}}
\altaffiltext{4}{Laboratoire de M\'et\'erologie Dynamique, Institut Pierre Simon Laplace, Paris, France}
\authors{James~W.~Head\altaffilmark{5}}
\altaffiltext{5}{Department of Geological Sciences, Brown University, Providence, RI 02912, USA}

\begin{abstract}
We use a 3D general circulation model to compare the primitive Martian hydrological cycle in ``warm and wet'' and ``cold and icy'' scenarios. In the ``warm and wet'' scenario, an anomalously high solar flux or intense greenhouse warming {artificially added to the climate model} are required to maintain warm conditions and an ice-free northern ocean. Precipitation shows strong surface variations, with high rates around Hellas basin and west of Tharsis but low rates around Margaritifer Sinus (where the observed valley network drainage density is nonetheless high). In the ``cold and icy'' scenario, snow migration is a function of both obliquity and surface pressure, and limited episodic melting is possible through combinations of seasonal, volcanic and impact forcing. At surface pressures above those required to avoid atmospheric collapse ($\sim0.5$~bar) and moderate to high obliquity, snow is transported to the equatorial highland regions where the concentration of valley networks is highest. Snow accumulation in the Aeolis quadrangle is high, indicating an ice-free northern ocean is not required to supply water to Gale crater. At lower surface pressures and obliquities, both \ce{H2O} and \ce{CO2} are trapped as ice at the poles and the equatorial regions become extremely dry. 
{The valley network distribution is positively correlated with snow accumulation produced by the ``cold and icy'' simulation at 41.8$^\circ$ obliquity but uncorrelated with precipitation produced by the ``warm and wet'' simulation.} Because our simulations make specific predictions for precipitation patterns under different climate scenarios, they motivate future targeted geological studies. 
\end{abstract}

\begin{article}

\section{Background}\label{sec:intro}

Despite decades of research, deciphering the nature of Mars' early climate remains a huge challenge. Although Mars receives only 43\% of the solar flux incident on Earth, and the Sun's luminosity was likely 20-30\% lower 3-4~Ga, there is extensive evidence for aqueous alteration on Mars' late Noachian and early Hesperian terrain. This evidence includes dendritic valley networks (VNs) that are distributed widely across low to mid latitudes \citep{Carr:96,Mangold2004,Hynek2010}, open-basin lakes \citep{Fassett2008a}, in-situ observations of conglomerates \citep{Williams2013}, and spectroscopic observations of phyllosilicate and sulphate minerals \citep{Bibring2006,Mustard2008,Ehlmann2011}. 

All these features indicate a pervasive influence of liquid water on the early Martian surface. Nonetheless, key uncertainties in the nature of the early Martian surface environment remain.  These include the intensity and duration of warming episodes, and the extent to which the total surface \ce{H2O} and \ce{CO2} inventories were greater than today. Broadly speaking, proposed solutions to the problem can be divided into those that invoke long-term warm, wet conditions (e.g., \cite{Pollack1987,Craddock2002}), and those that assume the planet was mainly frozen, with aquifer discharge or episodic / seasonal melting of snow and ice deposits providing the necessary liquid water for fluvial erosion (e.g., \cite{Squyres1994,Toon2010,Wordsworth2013a}). 

Previously, we have shown that long-term warm, wet conditions on early Mars are not achieved even when the effects of clouds, dust and \ce{CO2} collision-induced absorption (CIA) are taken into account in 3D models \citep{Wordsworth2013a,Forget2013}, because of a combination of the faint young Sun effect \citep{Saga:72} and the condensation of \ce{CO2} at high pressures \citep{Kasting1991}. Instead, we have proposed that the migration of ice towards high altitude regions due to adiabatic surface cooling (the `icy highlands' effect) may have allowed the continual recharge of valley network sources after transient melting events in a predominately cold climate \citep{Wordsworth2013a}. 

\section{Motivation}

The spatial distribution of the ancient fluvial features on Mars is strongly heterogeneous. In some cases, such as the northern lowlands, the absence of surface evidence for liquid water is clearly a result of coverage by volcanic material and/or impact debris in the Hesperian and Amazonian epochs \citep{Tanaka1986,Head2002}. However, even in the Noachian highlands, the distribution of VNs \citep{Hynek2010}, open-basin lakes \citep{Fassett2008a} and phyllosilicates \citep{Ehlmann2011} is spatially inhomogeneous, with most fluvial features occurring at equatorial latitudes (north of 60$^\circ$~S). In addition, some equatorial regions such as Arabia and Noachis also exhibit low VN drainage densities, despite the fact that impact crater statistics indicate that the terrain is mainly Noachian \citep{Tanaka1986}.

In comparisons between climate models and the geologic evidence for early Mars, the focus is generally placed on the surface temperatures obtained, based on the argument that mean or transient temperatures above 0~$^\circ$C are required to explain the observed erosion. Equally important, but less explored, is the extent to which the \emph{spatial distribution} of fluvial features can be explained by climate models. The VN distribution has been comprehensively mapped in recent years \citep{Hynek2010}, so it constitutes a particularly important constraint on models of {the early Martian water cycle}. Here we begin by exploring a few of the most obvious explanations for why VNs are observed on some regions of Mars' ancient crust but not on others.

\begin{figure}[h]
	\begin{center}
		{\includegraphics[width=4in]{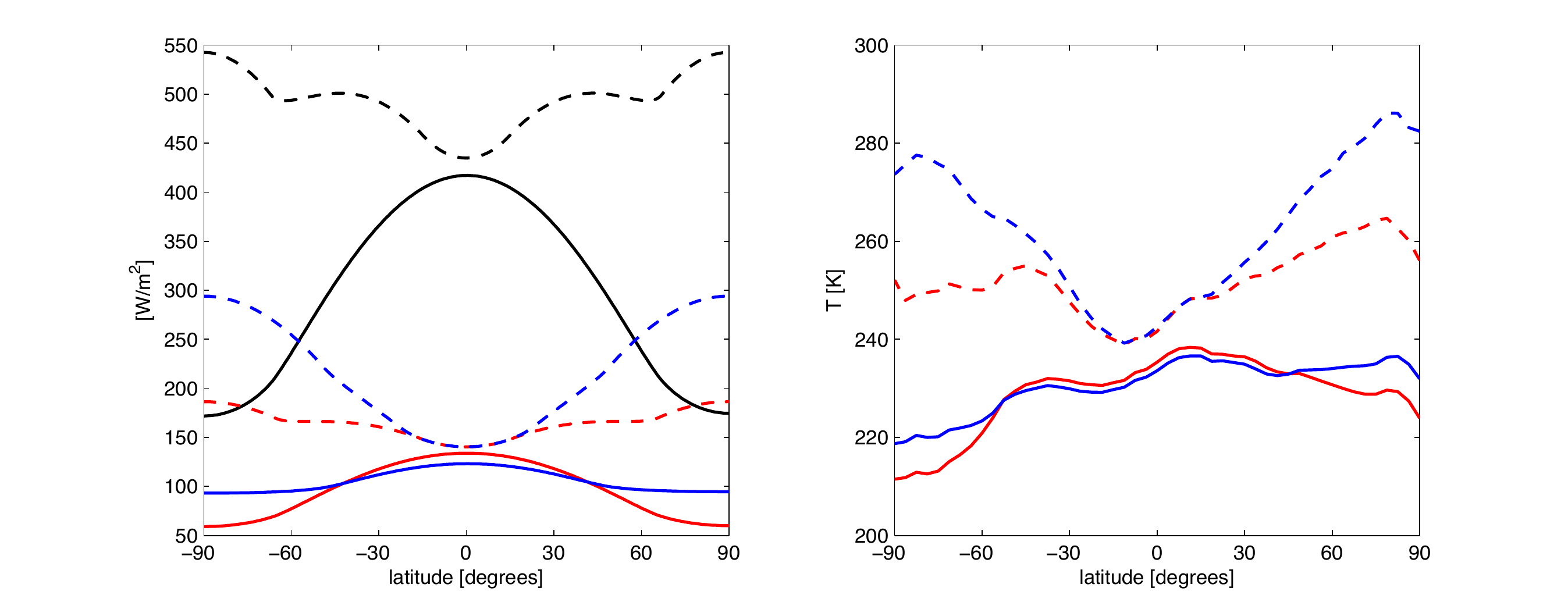}}
	\end{center}
	\caption{(left) Insolation vs. latitude for Earth (black) and early Mars with obliquity $25.0^\circ$ (red) and $41.8^\circ$ (blue).  Solid and dashed lines denote annual mean and annual maximum (diurnal mean) temperatures, respectively. For Earth the obliquity is $23.4^\circ$. In both cases an albedo of 0.2 and orbital eccentricity of 0.0 is assumed, while for early Mars the solar constant is taken to be 0.75 times the present-day value. (right) The corresponding annual mean (solid) and maximum (dashed) surface temperatures derived from dry early Mars 3D climate simulations with 1~bar surface pressure.}
\label{fig:insol}
\end{figure}

One seemingly obvious explanation for the predominance of VNs at equatorial latitudes is that early Mars was warmest at the equator, like Earth today. In an analogous climate to Earth's with abundant water sources, this would increase either precipitation or the frequency of melting events at the equator, and hence the erosion rate. However, this explanation for the VN distribution is problematic, for two reasons. First, if early Mars had a thicker \ce{CO2} atmosphere of 1-2~bar \citep{Phillips2001}, equator-pole heat transport by the atmosphere would have been much more effective than it is today, which would have decreased the latitudinal temperature gradient. Second, Mars' obliquity evolves chaotically due to secular orbital perturbations. At obliquity $\phi=41.8^\circ$ (the most probable value predicted over timescales greater than a few Gy \citep{Laskar2004}), the annual mean insolation for Mars 3.8~Gy at the poles is only 30.0~W~m$^{-2}$ less than that at the equator, while the annual maximum (diurnal mean) insolation is 153.6~W~m$^{-2}$ \emph{greater} (Figure~\ref{fig:insol}). Both these effects argue against a warmer equatorial climate as the reason for the observed valley network distribution.

What about alternative explanations? Another possibility is that differences in preservation dominate the spatial distribution, even on Noachian terrain. Difficulties in the detection of open-basin lakes poleward of $60^\circ$S has been noted \citep{Fassett2008a} and hypothesised to be a result of dust mantling at high southern latitudes \citep{Soderblom1973}. The presence of pedestal craters  poleward of $40^\circ$S also suggests that modification of high latitude terrain by ice in the Amazonian may have been significant \citep{Kadish2009}. However, because the Martian obliquity varies chaotically on geological timescales, even in the Amazonian water ice should have been transported from the poles to equatorial regions and back again many times over \citep{Madeleine2009}. If ice mantling has destroyed VNs that formed south of $60^\circ$S, therefore, it is not clear why it has not also destroyed them at equatorial latitudes.

The third and most interesting possibility is that at least some of the VN spatial heterogeneity is a result of \ce{H2O} supply limitations, due to regional differences in Noachian precipitation and/or sublimation/evaporation rates. If this is the case, it may provide clues as to the nature of the early Martian hydrological cycle, and hence the climate. To investigate this idea, we have performed a range of 3D GCM simulations under both cold and warm conditions.

\section{Method}\label{sec:method}

To compare various climate scenarios without constraints based on the specific combination of greenhouse gases available or the faintness of the young Sun, we assumed a \ce{CO2}-dominated atmosphere as in \cite{Wordsworth2013a} but studied the effects of a) progressively increasing the solar flux $F$ from the value of 75\% present-day predicted for $\sim$3.8~Ga from standard solar models and  b) adding a gray infrared absorption coefficient $\kappa$. We emphasize that such simulations are nominally unphysical: a significantly brighter young Sun is possible, but conflicts with both theoretical models of stellar evolution on the main sequence and observations of nearby Sun-like stars (e.g., \cite{MintonMalhotra2007}). Intense long-term greenhouse warming by gaseous absorption or infrared cloud scattering in a dense \ce{CO2} atmosphere has been previously proposed \citep{Pollack1987,Forget1997}, but neither mechanism currently appears physically viable \citep{Kasting1991,Wordsworth2013a,Forget2013}. Most recently, a \ce{H2}-\ce{CO2} atmosphere to create a long-term warm and wet early Mars has been suggested \citep{Ramirez2014}, following a similar proposal for \ce{H2}-\ce{N2} warming on the early Earth \citep{Wordsworth2013c}. This requires low \ce{H2} escape rates combined with very high combined \ce{H2} and \ce{CO2} outgassing in the late Noachian, which appears petrologically difficult to achieve \citep{HirschmannWithers2008}. Nonetheless, because the late Noachian geomorphology is often taken as evidence for a warmer, wetter climate, performing `empirical' 3D simulations of a warm, wet early Mars should still lead us to new insights.

For simplicity, we focus on two end-member cases here. In the first ``cold and icy'' case, a solar flux of $F=0.75F_0=441.1$~W~m$^{-2}$ is used as in \cite{Wordsworth2013a} to match Mars' received flux approximately 3.8~Ga \citep{Gough1981} and no additional gray-gas atmospheric absorption is included. 
In the second ``warm and wet'' case, $F$ is chosen to yield global mean temperatures sufficient for liquid \ce{H2O} precipitation (rain) in the southern highlands, and a northern ocean is assumed to be present. {In both cases,  the standard obliquity was taken to be $41.8^\circ$ but values ranging from $10^\circ-55^\circ$ were studied}. Key model parameters are given in Tables~\ref{tab:params1} and ~\ref{tab:params2}, and further details of the model setup are described below.

\subsection{Equilibrium temperature curves and valley network drainage density plot}

To produce the equilibrium temperature curves in Fig.~\ref{fig:insol}, we assumed a constant planetary albedo $A=0.2$ and calculated net insolation $F(L_s,\lambda)$ as a function of season angle $L_s$ and latitude $\lambda$ using standard expressions (e.g., \cite{Pierrehumbert2011BOOK}) and neglecting the diurnal cycle. The valley network density plot was calculated using the Line Density tool in the geographical information system software package ArcMap. Network density was calculated by considering a radius of 200~km around each pixel, summing the lengths of all the valley networks contained within it, and dividing this number by the area of the circle. Valley network outlines were taken from \cite{Hynek2010}, and geologic units were taken from \cite{Scott1986}, \cite{Greeley1987} and \cite{Tanaka1987}.

\subsection{General model features}
To produce the 3D simulation results we used the LMD generic climate model \citep{Wordsworth2010b}. Briefly, the model consists of a finite-difference enstrophy-conserving dynamical core that solves the meteorological primitive equations \citep{Sadourny1975} coupled to physical parameterizations for the planetary boundary layer and atmospheric radiative transfer. The latter is solved using a two-stream approach \citep{Toon1989} with optical data derived from line-by-line calculations \citep{Rothman2009} and interpolated collision-induced absorption data \citep{Clough1989,Gruszka1998,Baranov2004,Wordsworth2010}. We use the same spectral resolution as in \cite{Wordsworth2013a} {(32 infrared bands and 36 visible bands)} but an increased spatial resolution of $64\times48\times18$  {(longitude $\times$ latitude $\times$ altitude}). 

As before, the model contains three tracer species: \ce{CO2} ice, \ce{H2O} ice/liquid and H$_2$O vapour. Local mean CO$_2$ and H$_2$O cloud particle sizes were determined from the amount of condensed material and the number density of cloud condensation nuclei $[CCN]$, which was set at $10^5$~kg$^{-1}$ for \ce{CO2} ice, $5\times10^5$~kg$^{-1}$ for \ce{H2O} ice and $10^7$~kg$^{-1}$ for liquid \ce{H2O} droplets. The value of {$[CCN]$ is extremely difficult to constrain for paleoclimate applications. The values we have chosen for \ce{H2O} are simply best estimates based on comparison with observed values on Earth \citep{Hudson2002}. 
The dependence of climate on the assumed value of $[CCN]$ for both \ce{CO2} and \ce{H2O} are studied in detail in \cite{Wordsworth2013a} and \cite{Forget2013}. } 

 Convective processes and cloud microphysics were modeled in the same way as previously, with the exception of \ce{H2O} precipitation. This  was modeled using the parametrization of \cite{Boucher1995}, with the terminal velocity of raindrops calculated assuming a dynamic atmospheric viscosity and gravity appropriate for \ce{CO2} and Mars, respectively. The Boucher~et.~al. scheme is more physically justified than the threshold approach we used in \cite{Wordsworth2013a}; nonetheless in sensitivity tests (Figure~\ref{fig:warm_wet_rain}b) we found that the two schemes produced comparable time-averaged results.

\subsection{Warm and wet simulations}

In the warm and wet simulations, a 1~bar average surface pressure was assumed. Estimates of the maximum atmospheric \ce{CO2} pressure during the Noachian are of order 1-2~bar \citep{Phillips2001,Kite2014}, with one study \citep{Grott2011} pointing to lower values ($\sim 0.25$~bar) based on outgassing models that incorporate the lower estimated oxygen fugacity of typical Martian magmas \citep{HirschmannWithers2008}.

\begin{figure}[h]
	\begin{center}
		{\includegraphics[width=3.25in]{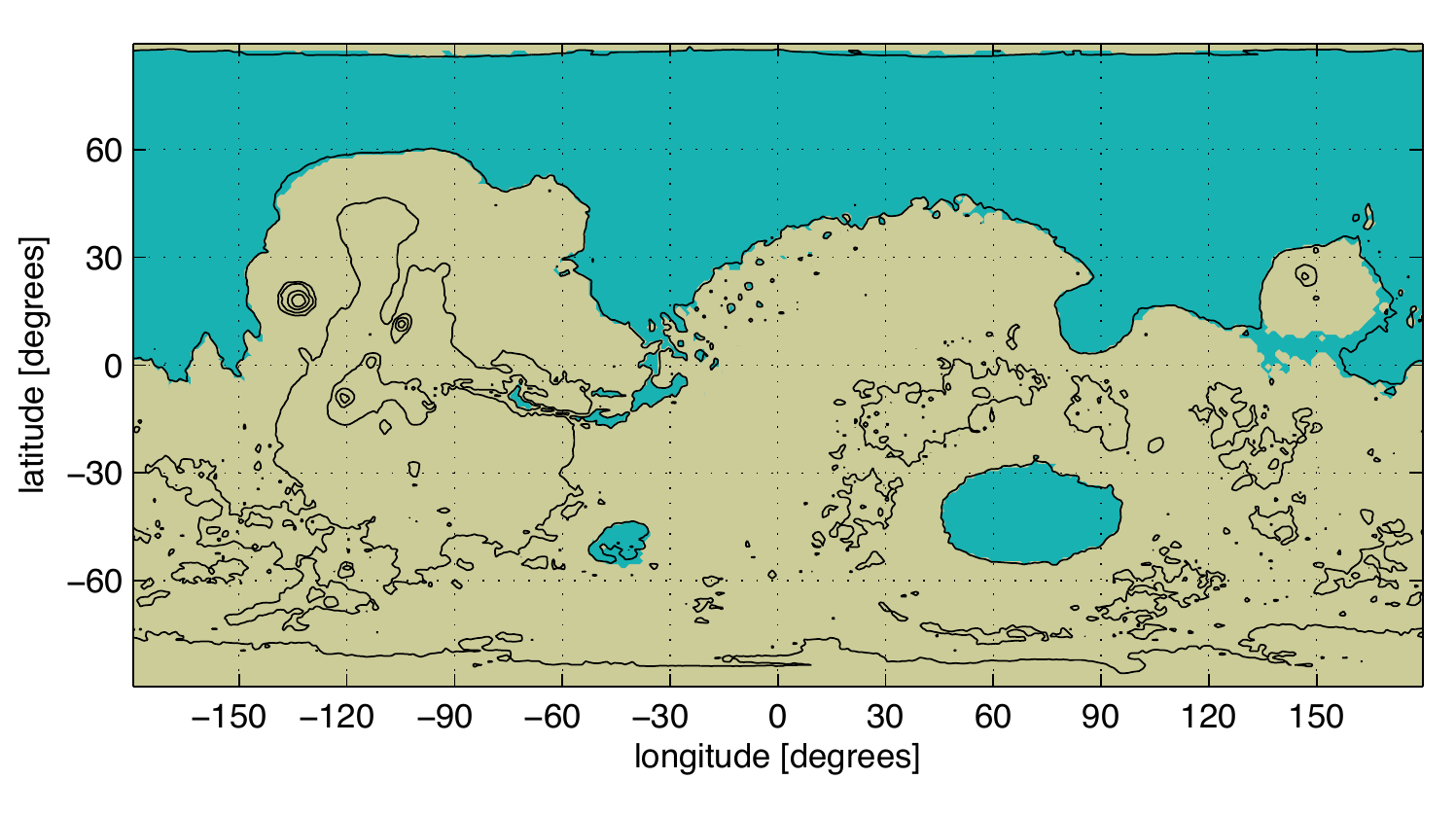}}
	\end{center}
	\caption{Plot of the surface water distribution in the warm, wet simulations. Brown and blue shading represents land and ocean, respectively.}
\label{fig:north_ocean}
\end{figure}

The topography was adjusted so that the minimum altitude was -2.54~km, {corresponding to the putative northern ocean shoreline based on delta deposit locations from \cite{diAchille2010}}. All regions at this altitude were then defined as `ocean': surface albedo was set to 0.07 and an infinite water source was assumed (Figure~\ref{fig:north_ocean}). In addition, the thermal inertia was adjusted to 14500~tiu, corresponding to an ocean mixed layer depth of approximately 50~m \citep{Pierrehumbert2011BOOK}. Horizontal ocean heat transport was neglected. Including it would have required a dynamic ocean simulator, which was outside the scope of this study. However, at 1~bar atmospheric pressure the meridional atmospheric heat transport is already effective, suggesting that inclusion of ocean heat transport would not significantly alter the main conclusions. 

On land, runoff of liquid water was assumed to occur once the column density in a gridpoint exceeded $q_{max,surf}=150$~kg~m$^{-2}$, {while the soil dryness threshold was taken to be 75~kg~m$^{-2}$, corresponding to subsurface liquid water layers of 15~cm and 7.5~cm, respectively. These values were chosen based on Earth GCM modeling \citep{Manabe1969}, and are reasonable first estimates given our simplified hydrology scheme and lack of knowledge of late Noachian subsurface conditions. The sensitivity of the results to the runoff threshold are investigated in Section~\ref{subsec:warmsens}}. 

The ice-albedo feedback was neglected in the warm and wet simulations {by setting the albedo of snow/ice to the same value as that of rock (0.2)}.  Its inclusion would have increased the chance of significant glaciation occurring in the simulations for a given solar flux or atmospheric infrared opacity and hence made a warm, wet Mars even more difficult to achieve. An investigation of the stability of an open northern ocean on early Mars to runaway glaciation will be given in future work. 

Finally, for the idealised equatorial mountain simulations that were run to study the effect of Tharsis on the hydrological cycle, we used
\begin{equation}
\phi(\theta,\lambda) = 2\times10^3 g e^{-{(\chi\slash40^\circ)}^2}
\label{eq:idealtharsis}
\end{equation}
with $\chi^2 = (\theta+100^\circ)^2+\lambda^2$, and longitude $\theta$ and latitude $\lambda$ expressed in degrees. This describes a flat surface with a Gaussian height perturbation {of full width at half maximum $66.6^\circ$} centered at $0^\circ$~N and $100^\circ$~W.

\subsection{Cold and icy simulations}

In the cold and icy simulations, 0.6~bar average surface pressure was assumed. This value was chosen as the minimum necessary to ensure atmospheric stability against collapse of \ce{CO2} on the surface for a wide range of obliquities. In some simulations, \ce{CO2} ice appeared in specific regions on the surface all year round. However, secular increases in the total surface \ce{CO2} ice volume after multiple years were not observed. 

{The local stability of surface ice on a cold planet is controlled by two parameters: the accumulation rate and the sublimation rate. The sublimation rate can be calculated based only on the near-surface temperature, humidity, and wind speed. Calculating the accumulation rate, in contrast, requires a fully integrated (and computationally expensive) representation of the hydrological system. In \cite{Wordsworth2013a}, we handled this problem using an ice evolution algorithm. We ran the climate model with a full water cycle for several years, calculating the tendencies for sublimation and accumulation. We then multiplied them by a factor between 10 and 100 and restarted the process until the system approached an equilibrium state.}

{Here, the increased spatial resolution of our model makes even this method prohibitively expensive computationally. Instead, we calculated ice accumulation rates with a full water cycle in two scenarios with different ice initial conditions [ a) polar source regions and b) an \ce{H2O} source below -2.54 km (the `frozen ocean' scenario)]. We also computed annual average potential sublimation rates for a wide range of parameters in simulations without a water cycle.}

{We discovered that surface thermodynamics are far more important than the details of the large-scale circulation for governing the surface ice evolution in the model, leading to a situation where the spatial distribution of  potential sublimation rates is inversely correlated with that of snow accumulation rates. This means that we can identify places where ice will stabilize in the long term far more rapidly than is possible using an ice evolution algorithm. Maps of potential sublimation can thus be used as a diagnostic for investigating the spatial distribution of stable ice.}

Quantitatively, we define the potential sublimation as the maximum possible  \ce{H2O} mass flux from the surface to the atmosphere 
\begin{equation}
\mathcal S_{pot} \equiv -\frac{ C_D |\mathbf v|}{R_\ce{H2O}T_a}p_{sat}(T_s),\label{eq:pot_sub}
\end{equation}
assuming a) that the surface is an infinite source and b) that the atmospheric relative humidity is zero. 
Here $R_\ce{H2O}$ is the specific gas constant for \ce{H2O}. The quantities $|\mathbf v|$, $T_a$ and $T_s$ (the surface wind speed, the temperature of the lowest atmospheric layer and the surface temperature, respectively) were derived directly from the 3D model output and $p_{sat}$ (the saturation vapour pressure) was derived from the Clausius-Clayperon relation. 
{The drag coefficient $C_D$ is defined (in neutrally stable conditions) as $C_D = \left( \frac{\mathcal K}{ln[ z\slash{z_0}]}\right)^2$,
where  $\mathcal K = 0.4$ is the von K\'arm\'an constant, $z_0$ is the roughness height, and $z$ is the height of the first atmospheric layer. $C_D$ was set to a constant $2.75\times10^{-3}$ based on comparison with the GCM value, which itself was calculated using a fixed roughness length of $10^{-2}$~m.} 
Finally, we integrate  $\mathcal S_{pot}$ over one Martian year to give the theoretical maximum quantity of surface snow/ice that can be sublimated as a function of longitude and latitude. Our approach to calculating $\mathcal S_{pot}$ is accurate as long as the atmospheric thermodynamics is not strongly affected by the presence of \ce{H2O}. This is the case for a cold early Mars, which has a surface moist convection number $\mathcal M $ of much less than unity \citep{Wordsworth2013water}. 

To produce estimates of transient melting in the cold, icy scenario, we performed simulations where we started from the baseline cold climate state and allowed the system to evolve after altering various parameters. The {melting event} simulations that included dust ({Section~\ref{subsec:melting})} used the same method as in \cite{Forget2013}, with the total atmospheric optical depth at the reference wavelength (0.7~$\mu$m) set to 5. The value of $e=0.125$ in the high eccentricity simulations was chosen with reference to the orbital evolution calculations of \cite{Laskar2004}. The increase of solar flux from the baseline value of $0.75F_0$ to $0.8F_0$ in some of the simulations is justified given the uncertainty in the absolute timing of valley network formation as estimated from crater statistics \citep{Fassett2008}. For some simulations, the surface albedo was reduced. The rock albedo $A_r$ was taken to be that of basalt (0.1; plausibly the Martian surface was richer in mafic minerals in the Noachian era \citep{Mischna2013}), while the ice albedo $A_i$ was taken to be 0.3 ({an approximate lower limit for ice contaminated by dust and volcanic ash deposition; \cite{Conway1996}}). The simulations including \ce{SO2} radiative forcing used the same correlated-$k$ method as used for the \ce{CO2}-\ce{H2O} atmospheres, except that 10~ppmv of \ce{SO2} was assumed to be present at all temperature, pressure and \ce{H2O} mixing ratio values calculated. \ce{SO2} absorption spectra were obtained from the HITRAN database \citep{Rothman2009}, and the differences in line broadening due to the presence of \ce{CO2} as the dominant background gas were neglected. {Finally, the surface albedo was increased to $A_i$ and a thermal inertia appropriate to solid ice} (2000~tiu) was assumed in regions where more than $33$~kg~m$^{-2}$ (3.3~cm) of surface \ce{H2O} was present \citep{LeTreut1991}. {This implies a quite rapid increase of surface thermal inertia with snow deposition, which makes our results on the rate of summertime snowmelt in the cold scenario (Section~\ref{subsec:melting})  conservative}. We also neglected further surface radiative effects such as insulation by a surface snow layer or solid-state greenhouse warming by ice, {both of which could potentially cause some additional melting}.

\section{Results}

\subsection{Overview}\label{subsec:over}
\begin{figure}[h]
	\begin{center}
		{\includegraphics[width=3.25in]{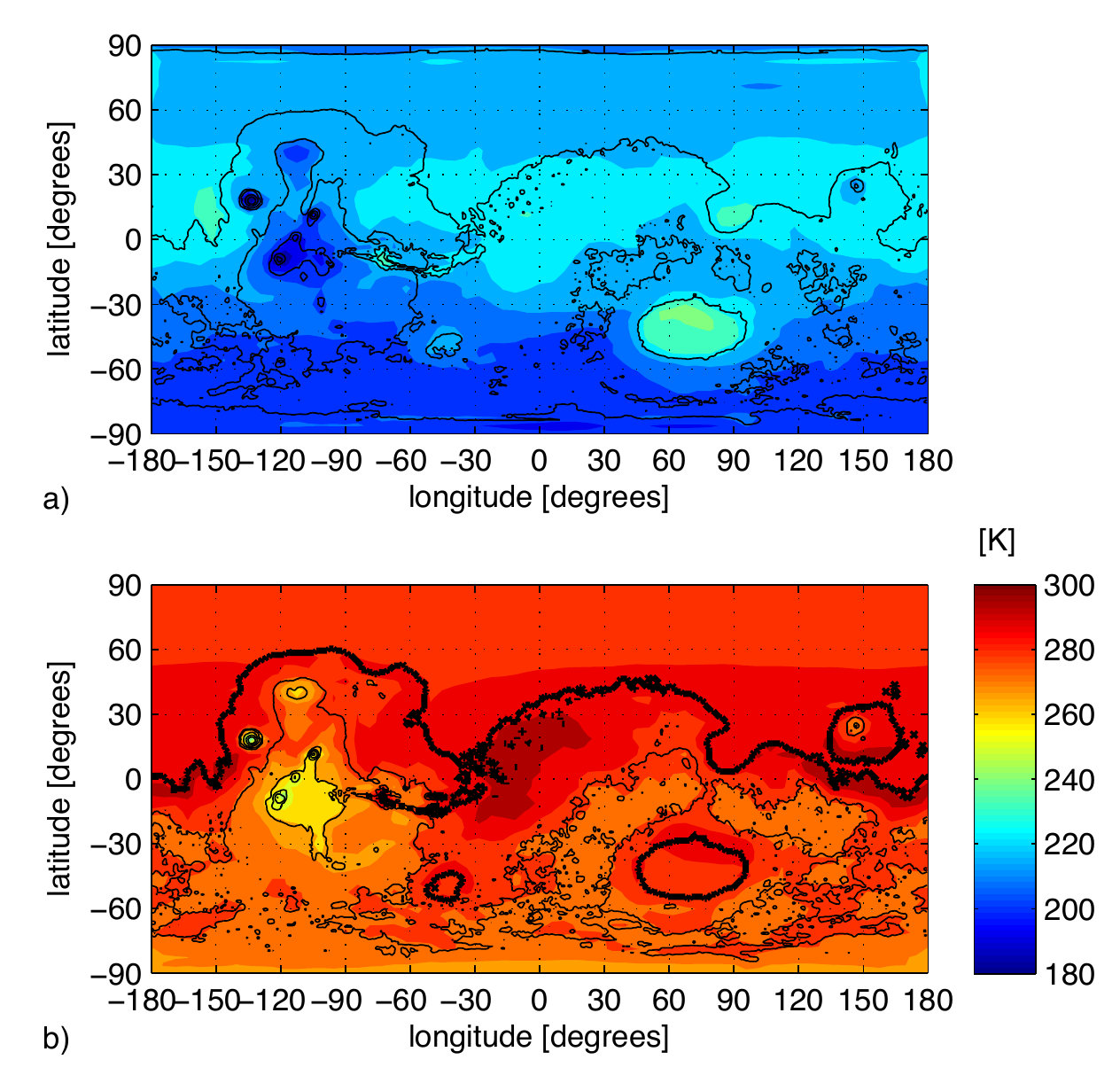}}
	\end{center}
	\caption{Plot of annual mean surface temperatures in a) the standard cold scenario and b) the standard warm, wet scenario. Obliquity is 41.8$^\circ$ in both cases. Global mean surface temperature is 225.5~K for a) and 282.9~K for b).}
\label{fig:Tsurfs}
\end{figure}

In our cold simulations, surface temperature is constrained by {a fully self-consistent climate model}, whereas in the warm, wet simulations, we can choose surface temperature by varying the solar flux and/or added gray opacity of the atmosphere. Fig.~\ref{fig:Tsurfs} shows annual mean surface temperatures in the two standard scenarios, given obliquity 41.8$^\circ$. In the cold case, surface pressure is 0.6~bar and solar flux is 0.75 times the present day value, while in the warm case the pressure is 1~bar and the solar flux is 1.3~times present day. Warm simulations where we used the best-estimate Noachian solar flux but added gray longwave opacity to the atmosphere showed broadly similar results (see Figure~\ref{fig:sens_warmwet}c).

As can be seen, in both cases adiabatic cooling influences surface temperature.  {In the cold, icy scenario,} the temperatures in the southern hemisphere are partly affected by seasonal condensation of \ce{CO2}. In the warm, wet scenario, horizontal temperature variations in the northern and Hellas basin oceans are low despite the absence of dynamical ocean heat transport, {because heat transport by the 1~bar atmosphere is efficient}. Global mean surface temperature is 225.5~K for the cold scenario and 282.9~K for the warm scenario.

Fig.~\ref{fig:warm_wet_rain} shows VN drainage densities computed from high-resolution topographic, visible and infrared data \citep{Hynek2010} alongside precipitation rates in the warm, wet scenario. For comparison, Fig.~\ref{fig:cold_icy_snow} shows snow/ice accumulation rates in the cold scenario. 
As can be seen, rainfall rates vary significantly with longitude as well as latitude. Over Noachian terrain, precipitation peaks occur in Arabia and around Hellas basin, including in the west and southeast where few VNs are observed. However, several regions with high VN drainage densities, such as Margaritifer Sinus, exhibit relatively low precipitation (approx. 10-20~times less than the peak value). In contrast, in the cold scenario (Fig.~\ref{fig:cold_icy_snow}a-b), surface ice/snow mainly migrates to equatorial latitudes, where the majority of the VNs are observed. The snow deposition is inhomogeneous spatially, with peak values of $\sim100$~kg~m$^{-2}$~(Mars~yr)$^{-1}$ in the regions of highest elevation declining to under $\sim0.1$~kg~m$^{-2}$~(Mars~yr)$^{-1}$ below $50^\circ$S. {Snow accumulation rates in the cold, icy case are orders of magnitude lower than precipitation rates in the warm case due to the exponential dependence of the evaporation/sublimation rate on temperature. This of course also means that the equilibration time of the surface hydrological cycle is much longer when Mars is cold \citep{Wordsworth2013a}.}

\subsection{Warm and wet scenario: Latitude dependence of precipitation}\label{subsec:swamp}

To understand the water cycle in our warm, wet simulations intuitively, we performed several idealized simulations. First, we calculated precipitation patterns in simulations with flat surface topography, infinite \ce{H2O} sources at every gridpoint, and entirely gray radiative transfer. Fig.~\ref{fig:swamp} shows the results of these `swamp planet' simulations for a range of orbital obliquities. 

{As can be seen, at all obliquities precipitation is highest near the equator in a narrow band of latitudes that varies with season. This band is the inter-tropical convergence zone (ITCZ). It is caused by regions of rapidly ascending air in the upward branches of the north- and southward Hadley cells \citep{Holton2013}. 
The second major feature of Fig.~\ref{fig:swamp} is that for $\phi=25^\circ$ or greater, significant precipitation occurs at all latitudes, although in a pattern that appears distinct from the ITCZ.}

{If instantaneous adjustment of the global mean flow to the solar forcing is assumed, moderate seasonal excursions of the ITCZ are possible \citep{Lindzen1988}. This explains the sinusoidal behavior of the precipitation at $\phi=10^\circ$ and $25^\circ$ in Fig.~\ref{fig:swamp}. }
However, when the latitude of peak solar forcing is too high, the double constraints of energy and angular momentum conservation cannot be met in the axisymmetric limit, and flow instability followed by eddy growth is expected in the winter hemisphere. In addition, the Hadley circulation in the summer hemisphere of seasonal axisymmetric flows becomes weak \citep{Lindzen1988}. Given this, the `messier' regions of high latitude precipitation in Fig.~\ref{fig:swamp} at moderate to high obliquity must therefore be due to horizontally localized moist convection and/or eddy processes. Interestingly, precipitation maxima at high latitudes have also been modeled for the methane cycle on present-day Titan \citep{Mitchell2006}, {although the physical processes at play in that case are somewhat different.}

The localized nature of the high-latitude precipitation means that it is particularly sensitive to the presence of nearby water sources. Indeed, in simulations where we assumed Hellas Basin was dry, much less rainfall south of $60^\circ$~S was observed (results not shown). Such simulations are likely physically inconsistent, however, because a large \ce{H2O} reservoir combined with continual precipitation and groundwater infiltration should ensure the presence of liquid water in low geopotential regions across the planet's surface \citep{Clifford2001}.

\subsection{Warm and wet scenario: Tharsis rain shadow}

Another striking feature of the warm, wet simulations is the low precipitation around Margaritifer Sinus. Sensitivity studies indicated that this effect was caused by the presence of the Tharsis bulge. In our model, predominantly westerly surface winds are orographically lifted over Tharsis, causing enhanced rainfall on its western flank and reduced rainfall to the east. 

To understand the nature of this `rain shadow' effect, we performed further calculations with simplified topography. The orientation of the effect was somewhat perplexing initially, because the annual mean equatorial surface wind direction was \emph{easterly} in the flat simulations (as on Earth). To understand why the rain shadow appeared on the east side of Tharsis, we first performed 3D simulations using the idealised Gaussian topography described previously. To {decrease model complexity and  gain greater} insight, we assumed a dry, pure 1~bar \ce{CO2} atmosphere, used non-scattering gray gas radiative transfer ($\kappa_{lw}=2\times10^{-4}, \kappa_{sw}=0$) and set obliquity to $0^\circ$. {The choice of zero obliquity was made to remove the effects of seasonality from the results.}

Fig.~\ref{fig:ideal_tharsis} shows the results in terms of the vertical and horizontal velocity at a mid-tropospheric $\sigma$-level in the model (approximately 0.7~bar). As can be seen, the vertical velocity distribution is zonally asymmetric at the equator, peaking on the west side of the topographic perturbation.  This behaviour is qualitatively similar to that found in a simpler quasi-linear model of the present-day Martian atmosphere (see \cite{Webster1977}, Fig.~10). The zonal wind is westerly at most latitudes, which fits the picture of a predominately eastward rain shadow.

Insight into a possible wave origin for the phenomenon can be gained by studying the classic Gill solution to the problem of equatorial waves forced by localised heating \citep{Gill1980}. Elevated topographic regions like Tharsis appear as heat sources to the atmosphere because of the same adiabatic cooling effect responsible for the icy highlands phenomenon in a cold climate: adiabatic cooling decreases the atmospheric temperature with altitude, but the solar energy input to the surface generally remains constant or increases with altitude.

To get a sense of the linear response of the equatorial $\beta$-plane system to simultaneous topographic and thermal forcing without a full eigenfunction analysis, in Fig.~\ref{fig:gill} we have plotted the $u(x,y)$-component of the (dimensionless) Gill solution for both Kelvin and Rossby modes
\begin{equation}
u = \frac 12 [q_0(x) + q_2(x)(y^2-3)]e^{-\frac14 y^2}
\end{equation}
at the equator ($y=0$), with $q_0(x)$ and $q_2(x)$ defined as in equations (4.2) and (4.7) of \cite{Gill1980}. We assume damping $\epsilon=0.05$ and a width of forcing region $L=2$. In Fig.~\ref{fig:gill}b, the equatorial vertical velocity response $w$ that would result if a topographic perturbation of the form $H(x) = e^{-k^2x^2}$ were present [i.e., $w \sim H'(x) u(x,0) = -2 k^2 x H(x) u(x,0)$ based on an assumption of incompressibility] is plotted. As can be seen, because of the zonal asymmetry of the Kelvin and Rossby wave responses, the vertical velocity can become strongly enhanced on the western side of the topographic perturbation.

Note that in simulations we performed where Tharsis was removed, rainfall rates in the Margaritifer Sinus region were found to be high (results not shown). However, evidence from magnetic field observations and geodynamics modeling indicates that the bulk of Tharsis was already in place by the period of peak VN formation \citep{Phillips2001,Fassett2011}. This suggests that erosion due to precipitation in a warm, wet climate is not the explanation for VN formation in Margaritifer Sinus.

\subsection{Warm and wet scenario: Sensitivity studies}\label{subsec:warmsens}

To test the robustness of our results, we also studied the sensitivity of our precipitation plots to variations in various poorly constrained model parameters. Results are shown in Figure~\ref{fig:sens_warmwet}. As can be seen, the same broad pattern is seen in all cases studied, with small differences on regional scales. Use of the simplified precipitation scheme from \cite{Wordsworth2013a} (Fig.~\ref{fig:sens_warmwet}b) causes an increase in the global mean value, mainly due to increases in the north and over Tharsis. Keeping the solar flux at 0.75~$F_0$ but adding gray longwave opacity to the atmosphere (Fig.~\ref{fig:sens_warmwet}c) decreases the global mean precipitation relative to the standard case, as expected from energetic considerations \citep{Pierrehumbert2011BOOK}. It also causes precipitation to become more localized in specific regions. Changing the runoff threshold or number of cloud condensation nuclei (Fig.~\ref{fig:sens_warmwet}d-e) also has no major effect. Finally, decreasing obliquity to $25^\circ$ slightly increases the amount of precipitation in the south and north and decreases it over Tharsis and Arabia Terra, but otherwise has little effect (Fig.~\ref{fig:sens_warmwet}f). 

\subsection{Cold and icy scenario: ice migration and annual potential sublimation}

Our high-resolution simulations yield broadly similar results to those reported in \cite{Wordsworth2013a}. Under higher atmospheric pressures than today, adiabatic cooling of the surface due to increased thermal coupling with the denser atmosphere causes surface temperatures to decrease with height on Mars \citep{Forget2013,Wordsworth2013a}. This causes the equatorial highlands to be among the coldest regions of the Martian surface, leading to enhanced stability of snow deposits there (the `icy highlands' effect \citep{Wordsworth2013a}).  

In Fig.~\ref{fig:cold_icy_snow}, net surface \ce{H2O} snow/ice accumulation in simulations with a water cycle is plotted along annual potential sublimation in simulation with fixed atmospheric relative humidity but no active water cycle. As can be seen, at the most probable obliquity 3.8~Ga from \cite{Laskar2004} ($\phi=41.8^\circ$), increased sublimation due to the high peak summertime temperatures  causes the regions between $\sim50^\circ$S and $\sim80^\circ$S to remain comparatively ice-free (Fig.~\ref{fig:cold_icy_snow}a) and snow accumulation is spatially correlated with the VN drainage density (see also Fig.~\ref{fig:VNcomp}). Interestingly, when a frozen ocean is assumed present below -2.54~km (Fig.~\ref{fig:cold_icy_snow}c), precipitation (snowfall) occurs over wider regions than in the standard cold case, including south of Hellas and Argyre Basins. Nonetheless, over multiple years the region of peak snow/ice accumulation remains the southern equatorial highlands. 

{As discussed in Section~\ref{sec:method}, the annual potential sublimation $\mathcal S_{pot}$ (Fig.~\ref{fig:cold_icy_snow}c) is closely correlated spatially with the snow accumulation rate. This demonstrates that surface thermodynamics is the key factor governing surface ice evolution in the model.
The \emph{magnitude} of $\mathcal S_{pot}$ is significantly greater than the snow accumulation rate. This is expected because in reality the atmospheric relative humidity is non-zero and the surface is dry in many regions. $\mathcal S_{pot}$ indicates the regions where ice will stabilize in the long term, but it gives only an approximate upper limit to the total rates of ice/snow transport. The close correlation of $\mathcal S_{pot}$ with the snow accumulation rate in Figs.~\ref{fig:cold_icy_snow}a-b nonetheless validates its use as a diagnostic for investigating the spatial distribution of stable ice in other situations.}

{In Figure~\ref{fig:pot_evap}, we show the variation of $\mathcal S_{pot}$ with 
obliquity and surface pressure (Figure~\ref{fig:pot_evap}).} 
In general, the effect of obliquity on surface ice stability lessens as the pressure increases, because the atmosphere becomes more effective at transporting heat across the surface (Figure~\ref{fig:pot_evap}f). At 0.6~bar, the effect of obliquity is still fairly important, with lower values causing lower ice stability at the equator in general. At $\phi=25^\circ$ (Figure~\ref{fig:pot_evap}b) $\mathcal S_{pot}$ is  lower (and hence the ice is more stable) in the equatorial highland regions than on most of the rest of the surface, although the main cold traps have become the south pole and peak of the Tharsis bulge. At $\phi=10^\circ$ (Figure~\ref{fig:pot_evap}a), both the north and south poles are cold traps and water ice is driven away from equatorial regions. Indeed, our simulations showed that at this obliquity \ce{CO2} itself is unstable to collapse at the poles, even for a surface pressure as high as 0.6~bar. 

Transient melting of equatorial \ce{H2O} under a moderately dense or thin \ce{CO2} atmosphere at low obliquity is therefore only possible if ice migrates more slowly than obliquity varies. Variations in the Martian obliquity and eccentricity occur on 100,000~yr  timescales \citep{Laskar2004}. Ice migration rates are difficult to constrain because of the nonlinear dependence of sublimation rate on surface temperature, but the mean values in our model for 0.6~bar \ce{CO2} are of order 1~kg~m$^{-2}$~(Mars~yr)$^{-1}$ or $\sim 1$~mm~(Mars~yr)$^{-1}$, leading to $\sim$10,000~Mars~yr for the transport of 10~m global averaged equivalent of \ce{H2O}. Unless the Noachian surface water inventory was orders of magnitude higher than the $\sim34$~m estimated to be present today, which may be unlikely \citep{Carr2014}, the equator would therefore have been mainly dry at low obliquity. Hence in the cold, icy scenario for early Mars, equatorial melting events should only occur when the obliquity is 25$^\circ$ or greater and/or atmospheric pressure is high.

\subsection{Cold and icy scenario: episodic melting events}\label{subsec:melting}

If early Mars was indeed mainly cold, some mechanisms must still have been responsible for episodic melting. Possible candidates include seasonal and diurnal effects \citep{Richardson2005,Wordsworth2013a}, positive radiative forcing from atmospheric dust and/or clouds \citep{Forget1997,Pierrehumbert1998,UrataToon2013}, orbital variations,  meteorite impacts \citep{Segura2008,Toon2010} and \ce{SO2}/\ce{H2S} emission from volcanism \citep{Postawko1986,Mischna2013,Halevy2014,Kerber2015}. 

Episodic impact events were an indisputable feature of the late Noachian environment, and they may also have had a significant effect on climate. Testing of the hypothesis that impacts caused melting sufficient to carve the valley networks requires 3D modeling of climate across extreme temperature ranges, which we plan to address in future work. However, it is worth noting that the \emph{stable} impact-induced runaway greenhouse atmospheres for early Mars proposed recently \citep{Segura2012} are highly unlikely. The conclusion of runaway bistability for early Mars is based on the assumption of radiative equilibrium in the low atmosphere, which requires unphysically high supersaturation of water vapour \citep{Nakajima1992}. Under more realistic assumptions for atmospheric relative humidity, the small decrease in outgoing longwave radiation with surface temperature past the peak value in a clear sky runaway greenhouse atmosphere is not sufficient for hysteresis under early Martian conditions \citep{Pierrehumbert2011BOOK}.

We have systematically studied all the other effects mentioned above (Fig.~\ref{fig:melting}; see also \cite{Wordsworth2013a,Forget2013,Kerber2015}). Alone, they are not sufficient to cause global warming sufficient to explain the observations. This is clear from Fig.~\ref{fig:melting}b, which shows the diurnal mean temperature vs. time averaged over the region $0^\circ$W--$120^\circ$~W, $10^\circ$N--$30^\circ$~S (see box in Fig.~\ref{fig:melting}a). The cyan line indicates the standard model, while the other colours indicate runs where effects were added separately [reduced surface albedo and increased atmospheric dust, blue; orbital eccentricity of 0.125 and solar constant of $F = 0.8F_0$, red; 10~ppmv atmospheric \ce{SO2}, green]. {We find significantly lower warming by \ce{SO2} than was reported in \cite{Halevy2014}, primarily because our 3D model accounts for the horizontal transport of heat by the atmosphere away from equatorial regions (see also \cite{Kerber2015}).}

The \emph{maximum} diurnal mean temperature for any of the studied cases is $-20^\circ$~C (for the increased eccentricity and $F = 0.8F_0$ run), implying very limited melting due to any single climate forcing mechanism. In combination, however, these effects can cause summertime mean temperatures to rise close to 0$^\circ$C in the valley network regions. This is shown by the black line in Fig.~\ref{fig:melting}b; the gray envelope indicates the maximum and minimum diurnal temperatures. 

{Figure~\ref{fig:melting_2} shows the average quantity of surface liquid water in the maximum forcing case as a function of time, assuming a melting point of 0$^\circ$C (black line) and  -10$^\circ$C (blue line). Uncertainty in the freezing point of early Martian ice due to the presence of solutes means that a melting threshold of -10$^\circ$C is  reasonable in a mainly cold scenario \citep{Fairen2009}. As can be seen peak melting reaches 5-7~kg~m$^{-2}$ in summertime in the briny case. This is still significantly below the 150~kg~m$^{-2}$ runoff threshold used in the warm and wet simulations, which itself was based on estimates from warm regions on Earth lacking subsurface permafrost. }

{\cite{Hoke2011} estimate that intermittent runoff rates of a few cm/day on timescales of order $10^5$ to $10^7$ years are necessary to explain the largest Noachian VNs. Our melting estimates are somewhat below these values, even if runoff is assumed to occur almost instantaneously after melting. Nonetheless, the exponential dependence of melting rate on temperature means that only a slight increase in forcing beyond that used to produce Fig.~\ref{fig:melting_2} would bring the model predictions into the necessary regime. Furthermore, the details of melting processes in marginally warm scenarios may still allow for significant fluvial erosion, as evidenced by geomorphic studies of the McMurdo Dry Valleys in Antarctica \citep{Head2014} and recent modeling of top-down melting from a late Noachian icesheet \citep{Fastook2015}. Incorporation of a more sophisticated hydrological scheme in the model in future incorporating e.g. hyporheic processes will allow these issues to be investigated in more detail.}

In this analysis, we do not claim to have uniquely identified the key warming processes in the late Noachian climate. Nonetheless, we have demonstrated that melting in a cold, icy scenario is possible using combinations of physically plausible mechanisms. In contrast, achieving {continuous} warm and wet conditions is much more difficult, as demonstrated by our simulations, which require an \emph{increased} solar flux compared to present day (in conflict with basic stellar physics) and/or intense greenhouse warming in addition to that provided by a 1-bar \ce{CO2} atmosphere.

\subsection{Spatial correlation of valley networks with rainfall and snow accumulation rates}

Because the model predictions for precipitation in the warm, wet regime and snow accumulation in the cold, icy regime are quite different, it is interesting to compare the spatial correlations with the Hynek~et~al. valley network drainage density in each case. To do this, we first filtered the data by excluding points on terrain that was not dated to the Noachian era. Polar areas and younger terrains were excluded from the analysis because absence of observable VNs does not necessarily indicate that they never formed in these regions.

\begin{figure}[h]
	\begin{center}
		{\includegraphics[width=4in]{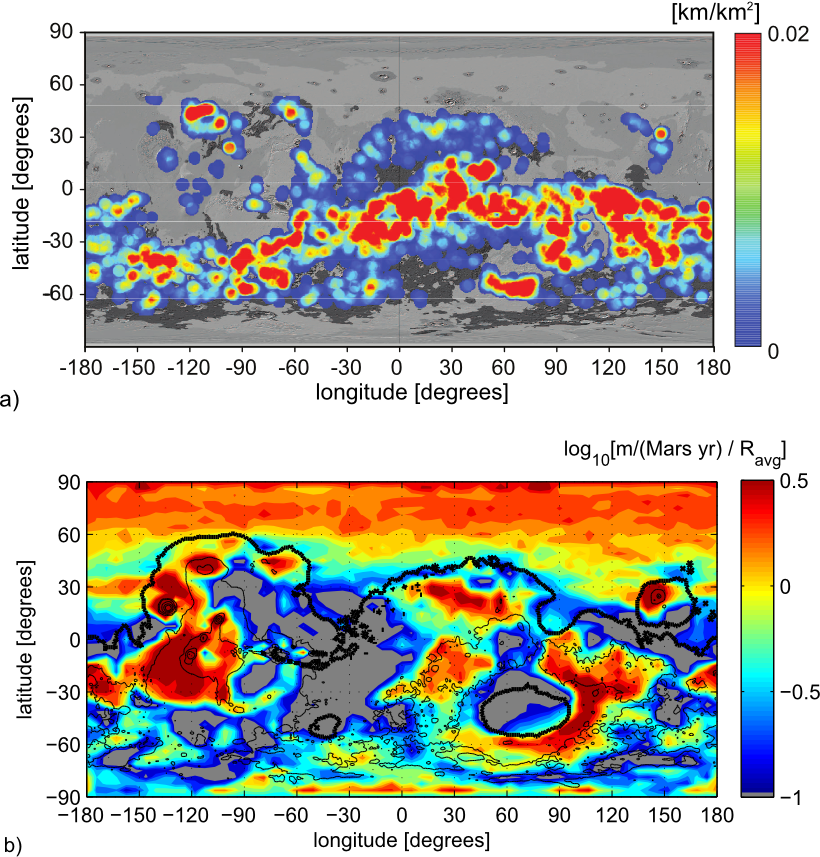}}
	\end{center}
	\caption{a) Surface VN drainage density (data derived from \cite{Hynek2010}), with shading of the background corresponding to terrain age (light, Amazonian; medium, Hesperian; dark, Noachian). b) Annual precipitation over 1~Martian year in the warm, wet simulation [global mean $R_{avg}=0.4$~m/(Mars yr)]. The solid black line indicates the ocean shoreline.  }
\label{fig:warm_wet_rain}
\end{figure}

\begin{figure}[h]
	\begin{center}
		{\includegraphics[width=4in]{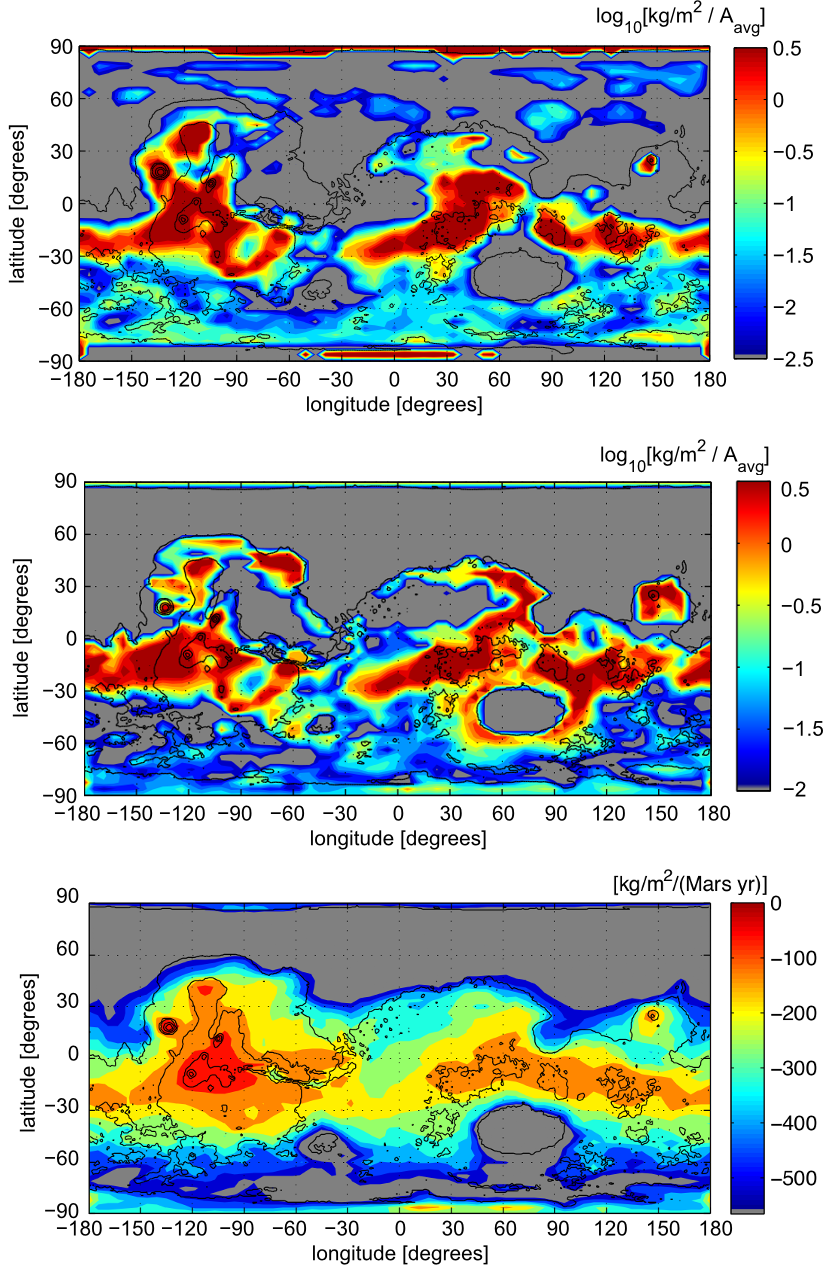}}
	\end{center}
	\caption{
	a) Surface snow/ice accumulation after 5~years in the default cold 3D climate simulation (obliquity $41.8^\circ$; pressure 0.6~bar) given a) \ce{H2O} sources at the poles; global mean $A_{avg}=5.3$~kg/m$^2$ and b)  an \ce{H2O} source and flat topography below -2.54~km; global mean value $A_{avg} = 3.2$~kg/m$^{2}$ (cold, wet scenario). c) Annual potential sublimation for a) and b).}
\label{fig:cold_icy_snow}
\end{figure}

\begin{figure}[h]
	\begin{center}
		{\includegraphics[width=5in]{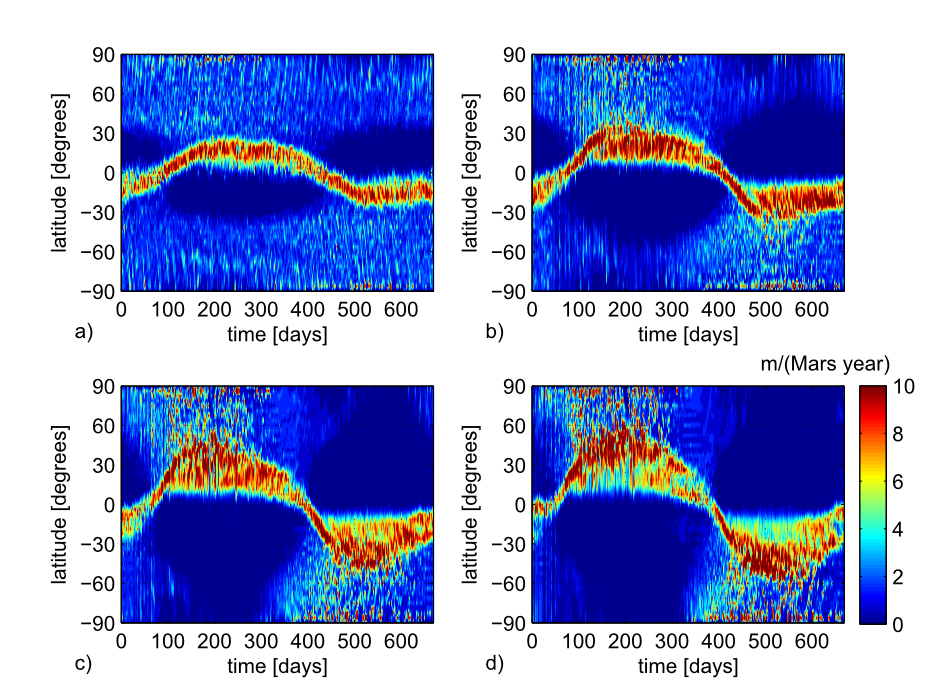}}
	\end{center}
	\caption{Annual precipitation (snow and rain) in idealised early Mars simulations with flat topography, non-scattering gray gas radiative transfer ($\kappa_{lw}=2\times10^{-4}$~m$^2$~kg$^{-1}$, $\kappa_{sw}=0.0$~m$^2$~kg$^{-1}$), uniform surface albedo of 0.2, solar flux $F=1.2F_0$, mean atmospheric pressure of 1~bar, no \ce{CO2} condensation, and an infinite surface \ce{H2O} reservoir at every gridpoint. a) obliquity $\phi=10^\circ$; b) $\phi=25^\circ$; c) $\phi=40^\circ$; d) $\phi=55^\circ$.}
\label{fig:swamp}
\end{figure}

\begin{figure}[h]
	\begin{center}
		{\includegraphics[width=5in]{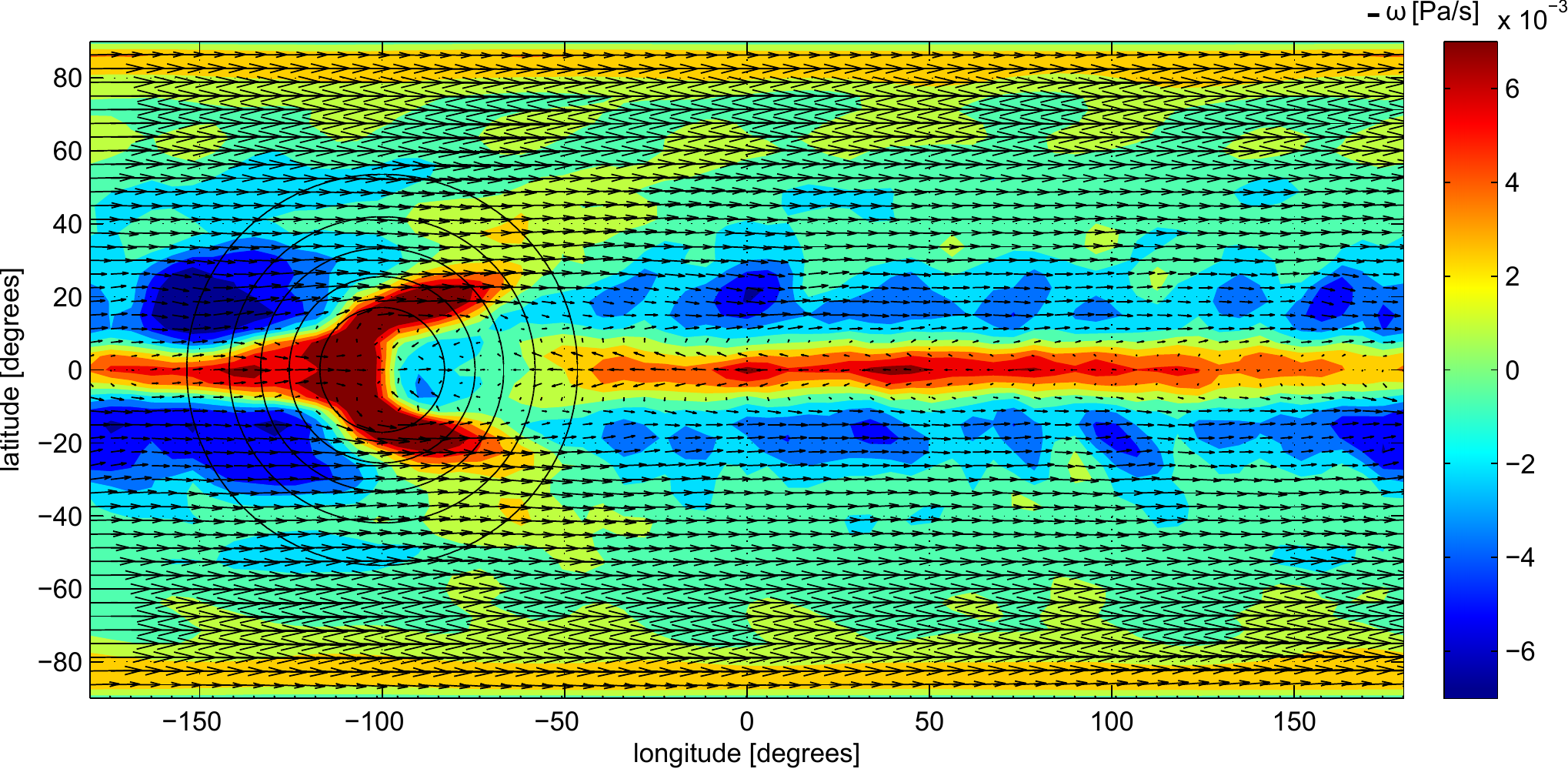}}
	\end{center}
	\caption{Annual mean vertical velocity (filled contours) and horizontal velocity (black arrows) at the 8th model level (approx. $p=0.7p_{surf}$) in an idealised dry, gray simulation with obliquity of 0$^\circ$ and topography represented by equation described in the main text (black solid lines).}
\label{fig:ideal_tharsis}
\end{figure}

\begin{figure}[h]
	\begin{center}
		{\includegraphics[width=4in]{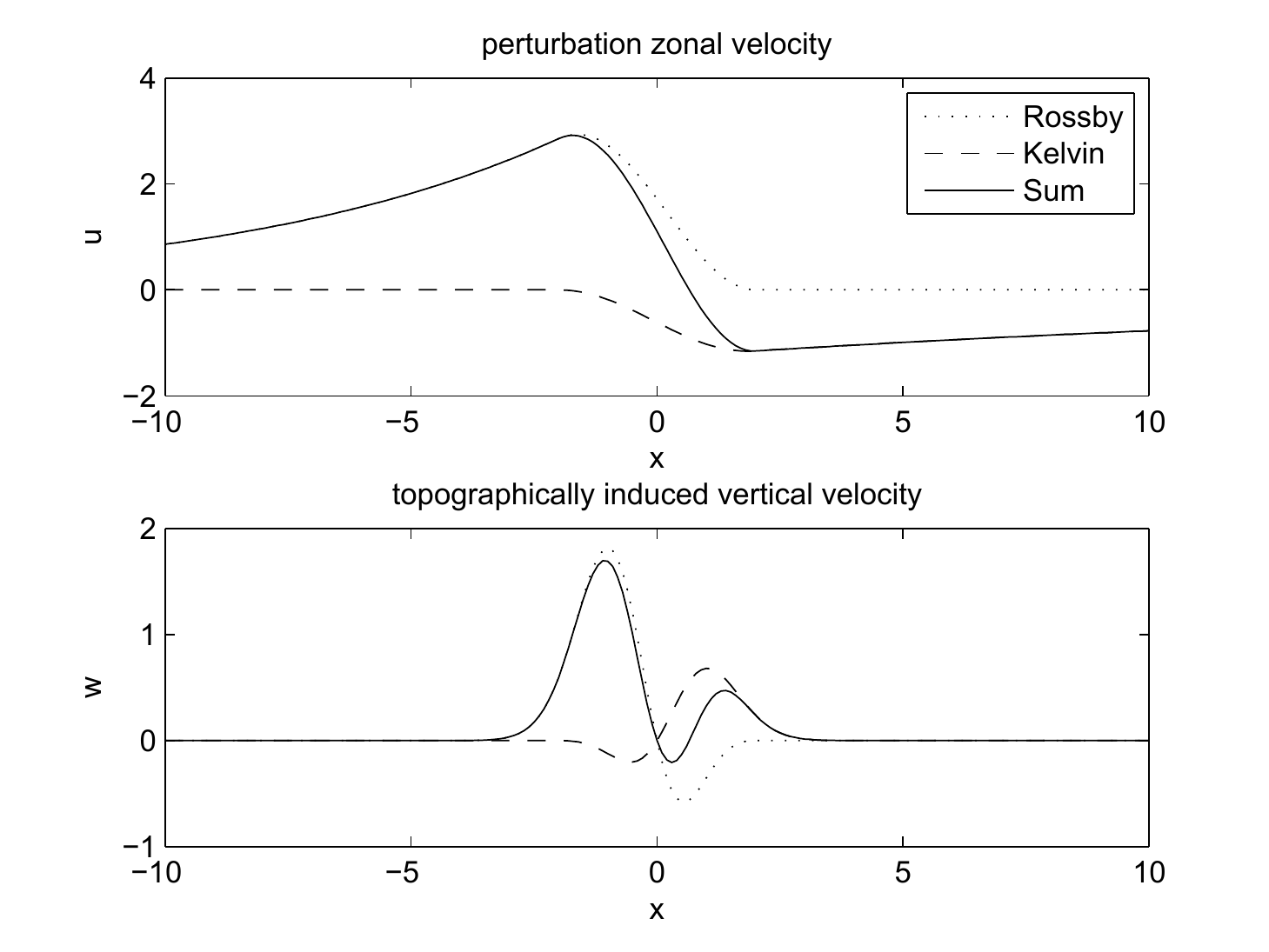}}
	\end{center}
	\caption{a) Zonal velocity in the Gill solution to the linear shallow water equations on an equatorial $\beta$-plane with local, spatially symmetric heating . b) Vertical velocity implied from a) due to mass conservation given the presence of a Gaussian topographic perturbation.}
\label{fig:gill}
\end{figure}

\begin{figure}[h]
	\begin{center}
		{\includegraphics[width=5in]{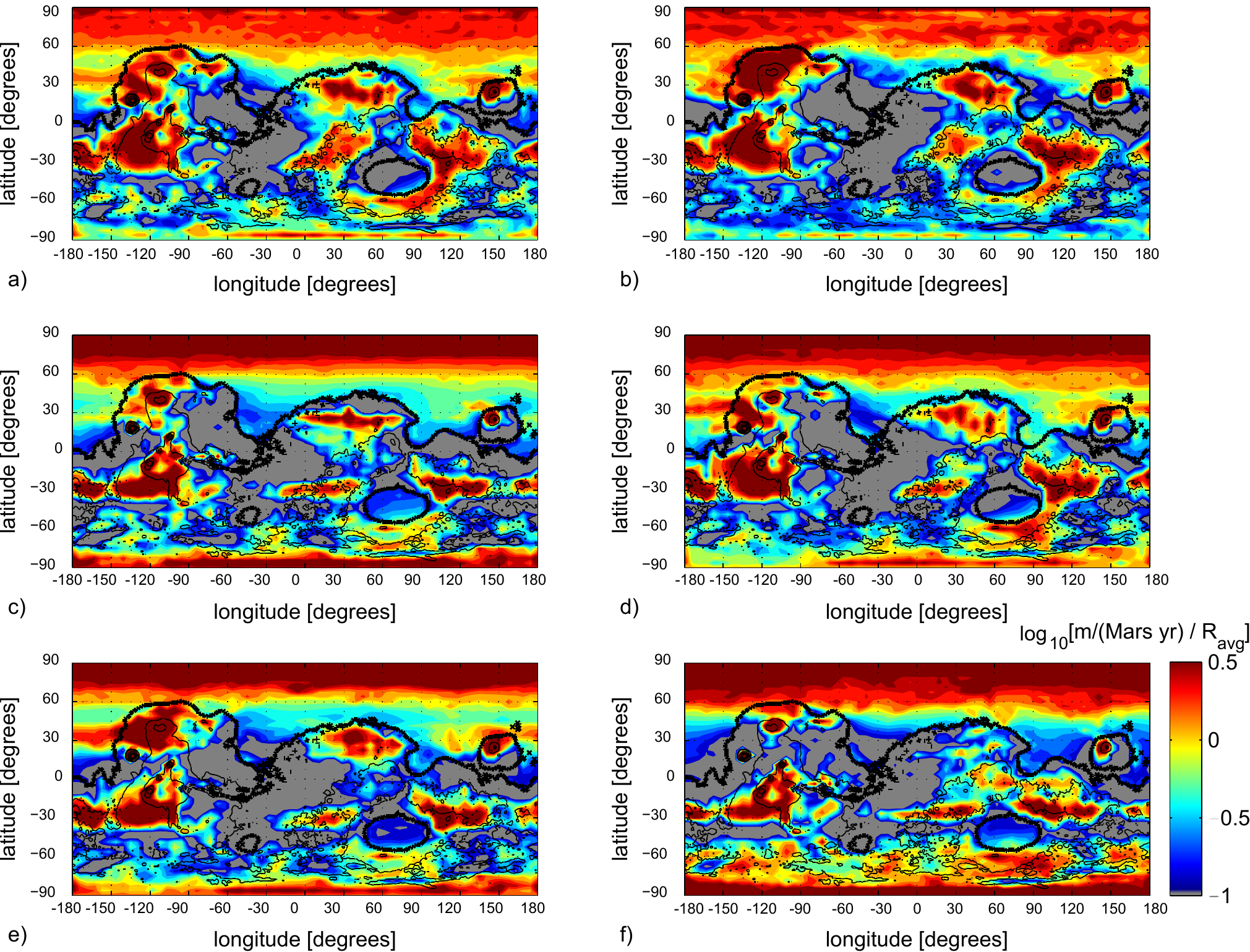}}
	\end{center}
	\caption{Annual precipitation in warm, wet runs {with ocean assumed as in Fig.~\ref{fig:north_ocean} and standard parameters as in Table~\ref{tab:params1}}. 
	a) Standard parameters as defined in the main text, b) a simplified threshold precipitation scheme as in Wordsworth~et~al. (2013), c) solar flux $0.75F_0$ but added gray gas atmospheric mass absorption coefficient $\kappa_{lw}=2\times10^{-4}$~m$^{2}/$kg, d) runoff threshold set to 1~kg~m$^{-2}$, e) variable cloud particle sizes with $[CCN]$ set to $1\times10^5$ for all particles, f) obliquity $\phi=25^\circ$. Global mean values ($R_{avg}$) are 0.41, 1.0, 0.31, 0.32, 0.65, 0.47~m/(Mars~yr) for a-f), respectively.}
\label{fig:sens_warmwet}
\end{figure}

\begin{figure}[h]
	\begin{center}
		{\includegraphics[width=5in]{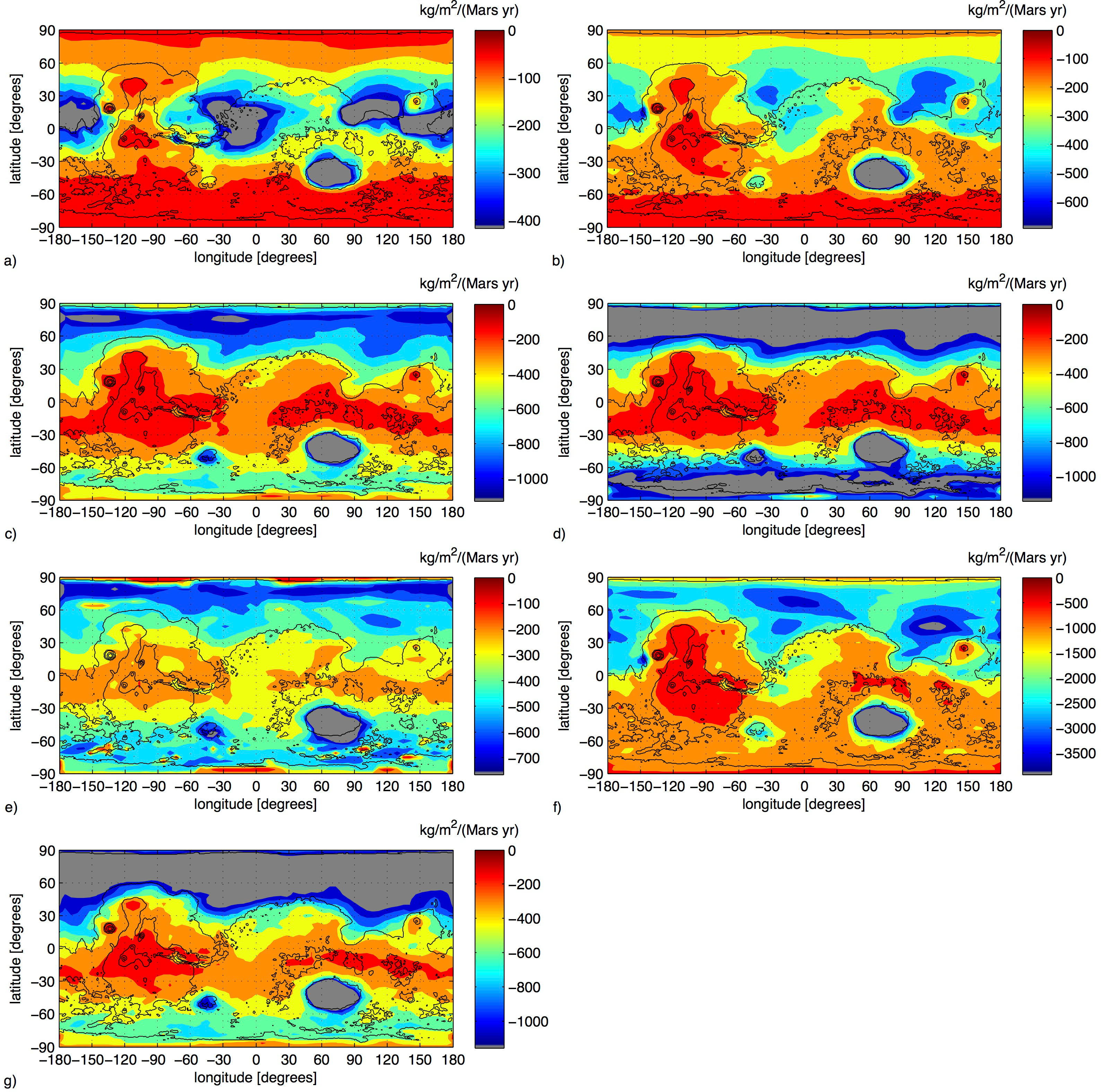}}
	\end{center}
	\caption{Annual potential sublimation in cold simulations with a) mean atmospheric pressure $p_\ce{CO2}=0.6$~bar, obliquity $\phi=10^\circ$  and eccentricity $e=0$, b), $p_\ce{CO2}=0.6$~bar, $\phi=25^\circ$  and $e=0$ c), $p_\ce{CO2}=0.6$~bar, $\phi=41.8^\circ$  and $e=0$,  d), $p_\ce{CO2}=0.6$~bar, $\phi=55^\circ$  and $e=0$  e), $p_\ce{CO2}=0.125$~bar, $\phi=41.8^\circ$  and $e=0$, f) $p_\ce{CO2}=2$~bar, $\phi=41.8^\circ$  and $e=0$  and g) $p_\ce{CO2}=0.6$~bar, $\phi=41.8^\circ$  and $e=0.125$.}
\label{fig:pot_evap}
\end{figure}

\begin{figure}[h]
	\begin{center}
		{\includegraphics[width=4in]{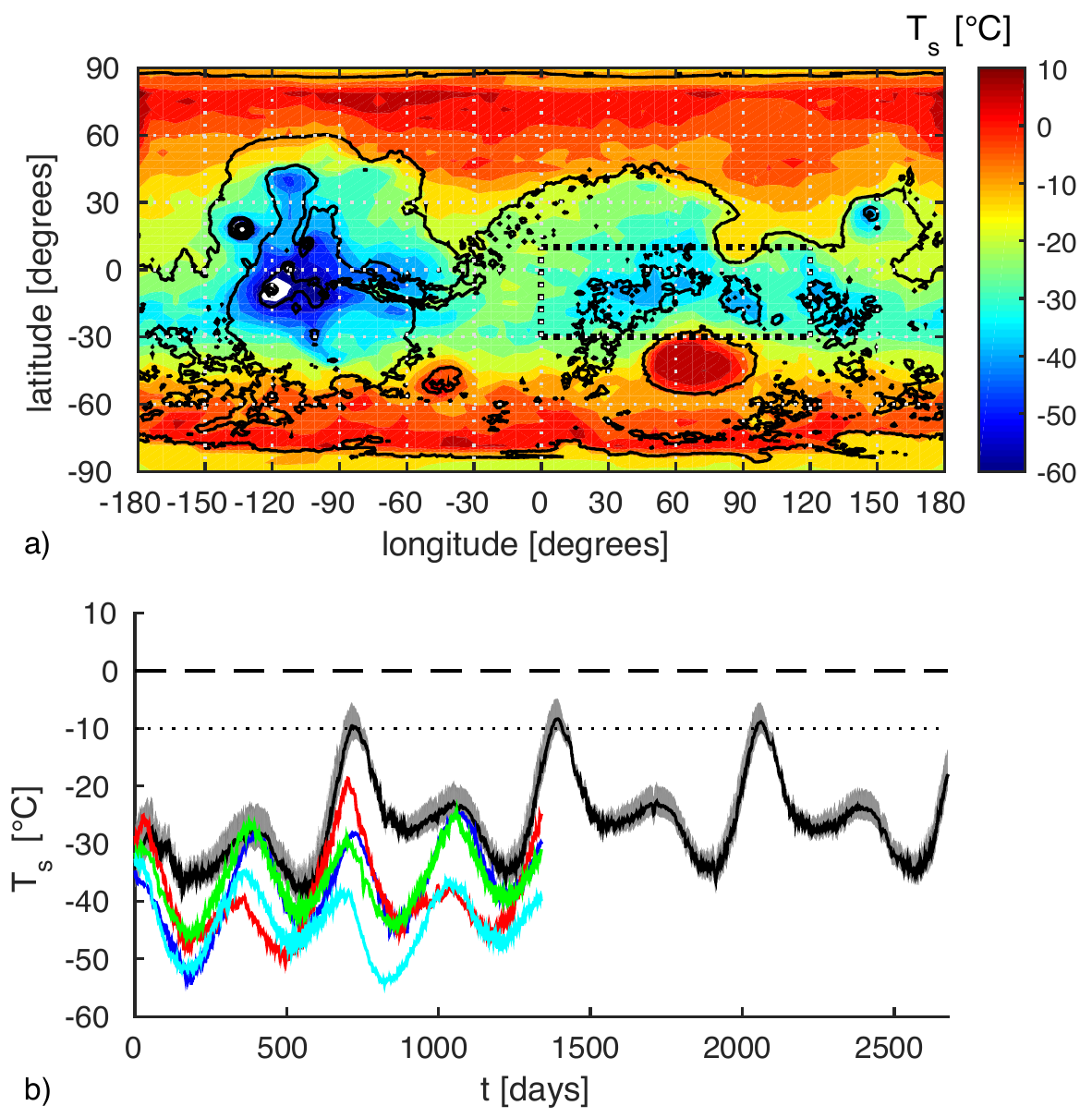}}
	\end{center}
	\caption{{Episodic warming in the cold, icy scenario.} a) Annual maximum diurnal mean surface temperature in the standard cold simulation. b) Diurnal mean temperature vs. time averaged over area denoted by the rectangle in the top panel for different forcing combinations [no forcing, cyan; reduced surface albedo and increased atmospheric dust, blue; orbital eccentricity of 0.125 and solar constant of $F=0.8F_0$, red; 10~ppmv atmospheric \ce{SO2}, green]. In black, a simulation with all of the previous effects is shown for four years simulation time. There, the extremal temperatures are indicated by the gray shading. The dashed line indicates the melting point of water, while the dotted line at -10$^\circ$~C indicates the point where up to 78\% of briny water would remain liquid according to \cite{Fairen2009}. }
	\label{fig:melting}
\end{figure}

\begin{figure}[h]
	\begin{center}
		{\includegraphics[width=5in]{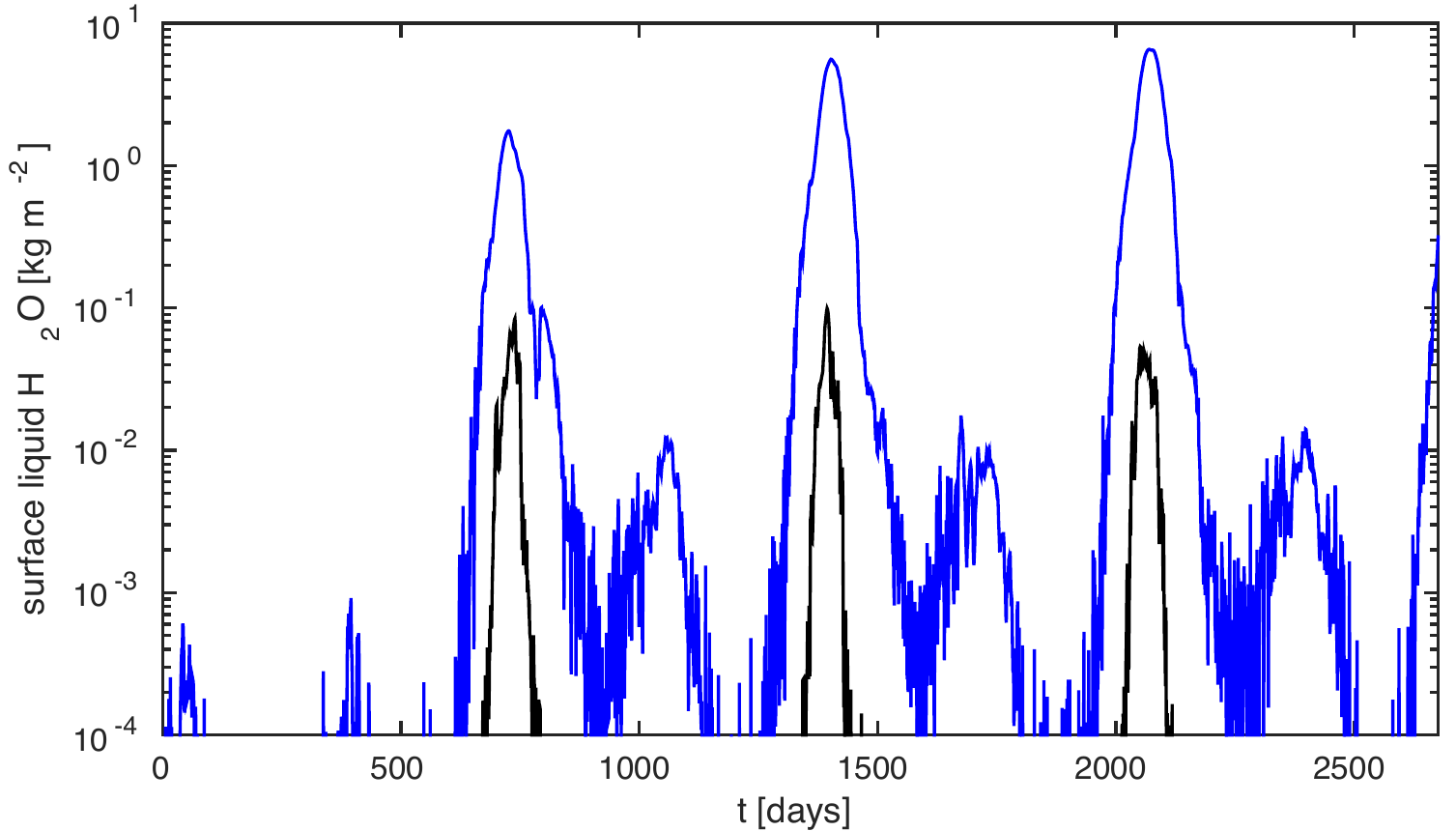}}
	\end{center}
	\caption{{Episodic melting in the cold, icy scenario. }Surface liquid water amount vs. time averaged over the same region as in Fig.~\ref{fig:melting} for the most extreme forcing case (black line in Fig.~\ref{fig:melting}). Here the black line shows surface liquid water in the case where 0$^\circ$~C is the melting point (fresh water), while the blue line shows the corresponding melt amount when -10$^\circ$~C is the melting point (brines).}
	\label{fig:melting_2}
\end{figure}

\begin{figure}[h]
	\begin{center}
		{\includegraphics[width=2in]{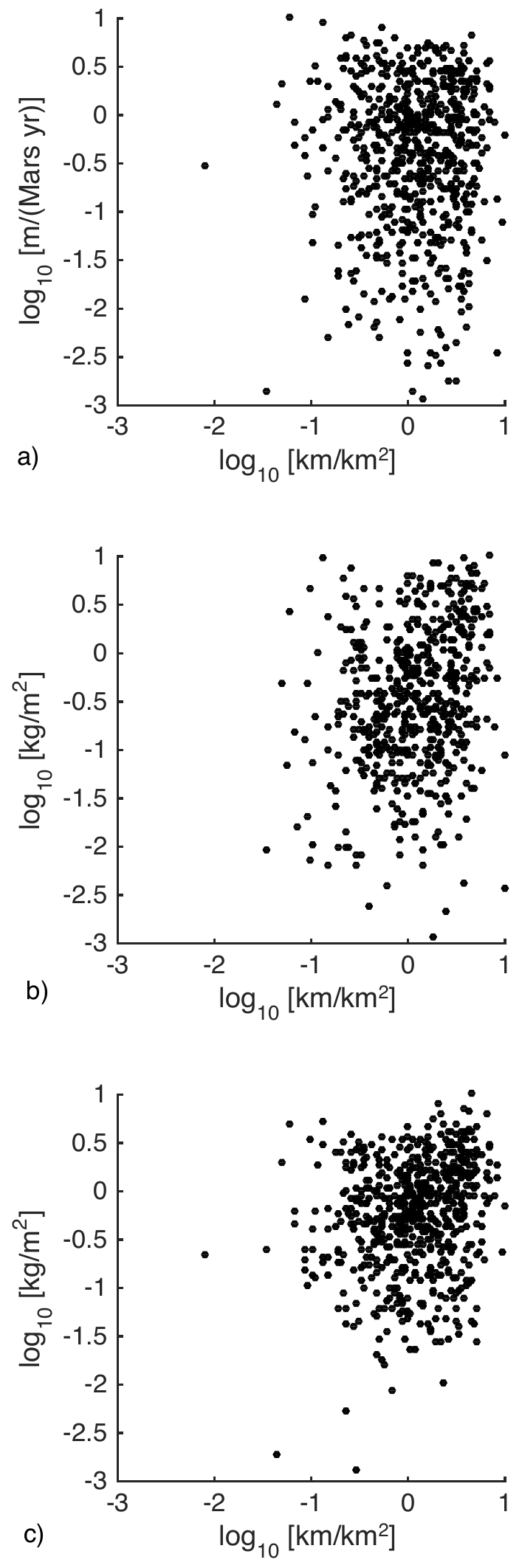}}
	\end{center}
	\caption{{Normalized logarithmic scatter plot of the VN drainage density from \cite{Hynek2010} vs. a) rainfall rate in the warm, wet scenario, b)  snow accumulation in the cold, icy scenario with polar water sources and c) snow accumulation in the cold, icy scenario with low altitude water sources. 
	From a-c), the correlation coefficients are -0.0033, 0.23 and 0.19, while the $p$-values are 0.93, $<0.001$ and $<0.001$, respectively. In all cases the total number of points is 644.}}
	\label{fig:VNcomp}
\end{figure}

Figure~\ref{fig:VNcomp} shows the resulting normalized logarithmic scatter plots of the VN drainage density (Fig.~\ref{fig:warm_wet_rain}a), a) precipitation in the standard warm and wet case (Fig.~\ref{fig:warm_wet_rain}b) and b-c) snow accumulation in the cold and icy cases (Fig.~\ref{fig:cold_icy_snow}a and b). For the warm and wet case, the correlation coefficient is -0.0033 and the $p$-value is 0.93, indicating the correlation is not statistically significant. For the two cold and icy cases, in contrast, the correlation coefficients are 0.23 and 0.19 while the $p$-values are both $<0.001$.

\begin{figure}[h]
	\begin{center}
		{\includegraphics[width=3in]{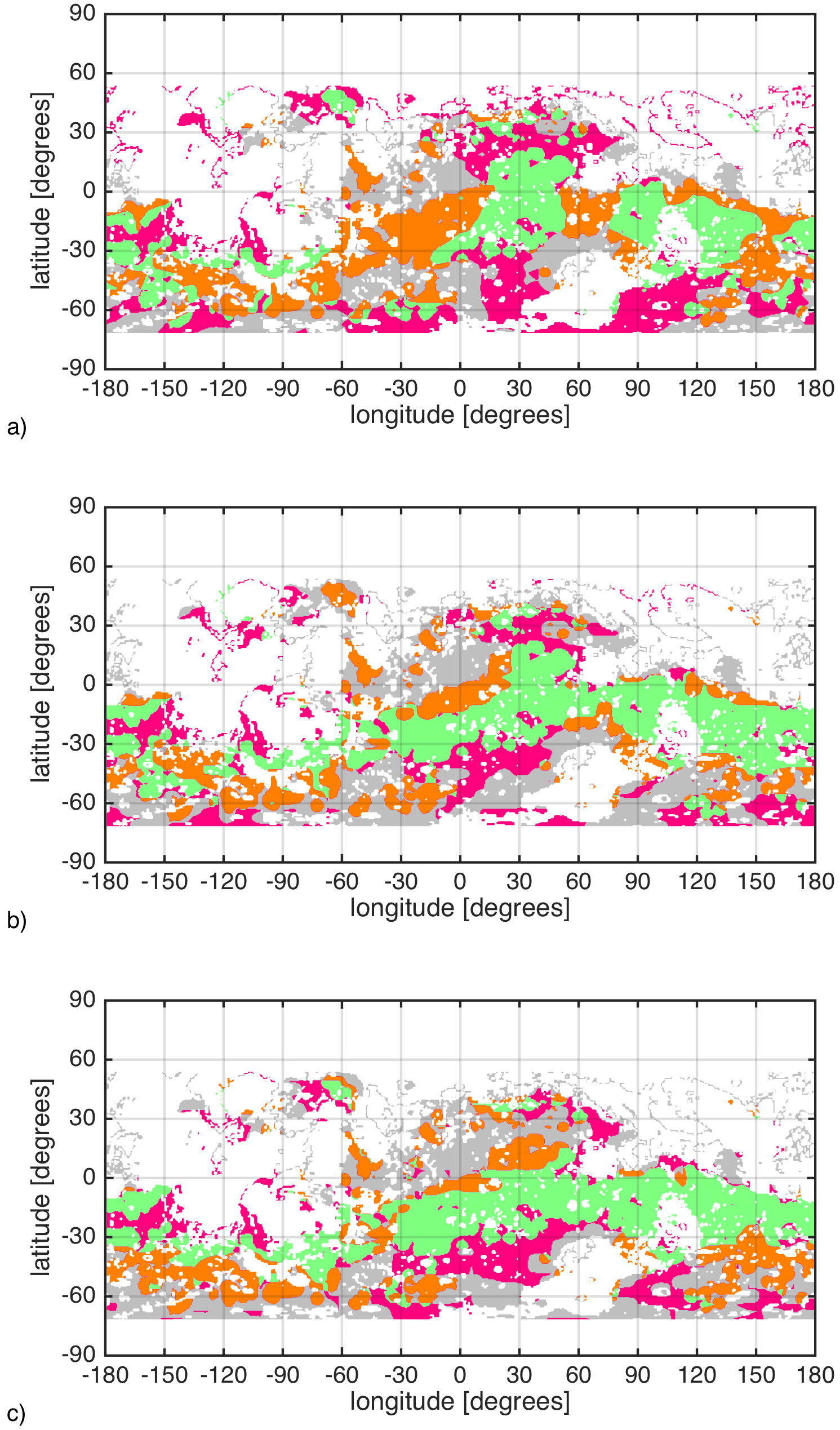}}
	\end{center}
	\caption{{Comparison of the VN drainage density from \cite{Hynek2010} with a) rainfall in the warm, wet scenario, b) snow accumulation in the cold, icy scenario with polar water sources and c) snow accumulation in the cold, icy scenario with low altitude water sources. Green indicates regions of agreement between VNs and rain/snow. Orange indicates significant presence of VNs but low rain/snowfall. Magenta indicates significant rain/snow but few VNs. Gray indicates low VN density and low rain/snowfall. Regions in white are where there was no data or the terrain is not Noachian age.}}
	\label{fig:VNcomp_2}
\end{figure}

The key reasons for the poor correlation in the warm, wet case appear to be a) the lack of precipitation around Margaritifer Sinus and b) the high precipitation on terrain south of 40$^\circ$S, where few VNs are present. This is clear from Fig.~\ref{fig:VNcomp_2}, which shows the regions of (dis)agreement between the VN drainage density and precipitation/snow accumulation maps across the surface.  Fig.~\ref{fig:VNcomp_2} was produced by creating a mask from the VN data with a dividing contour defined such that 50\% of the surface was denoted as containing VNs. Similar masks were then created from the warm and wet precipitation and cold and icy snow accumulation data, and for each plot the masks were superimposed to highlight regions of agreement and disagreement. 

Figure~\ref{fig:VNcomp_2} highlights the results discussed previously in Section~\ref{subsec:over}: in our model the warm and wet scenario disagrees most with the VN distribution in Margaritifer Sinus and south of Hellas Basin. The cold and icy scenario, in contrast, has closer agreement around the equator but still some disagreement on terrain south of 40$^\circ$S. In the cold and icy case, this discrepancy is somewhat artificial, because snow equilibration also varies significantly with obliquity. In the warm and wet case, the basic features of the precipitation map are much less sensitive to obliquity changes (Sections~\ref{subsec:swamp}-\ref{subsec:warmsens}).

The lack of correlation between warm and wet precipitation and the VN distribution when compared to the cold and icy result is striking. Because prediction of precipitation patterns using GCMs is non-trivial even for the present-day Earth, further studies in future using a range of convection schemes and model dynamical cores will be important. If the dynamical effects we have analyzed here turn out to be independent of all model details, this clearly will constitute a significant piece of evidence against a warm and wet origin for many of the valley networks.

\section{Discussion}

We have presented the first direct comparison of warm, wet and cold, icy early Mars scenarios in a 3D climate model. Because of their generality, 3D models allow hypotheses for early Mars to be compared with the geological evidence and tested for internal consistency to a far greater extent than is possible with 1D radiative-convective or 2D energy-balance models.

In our warm and wet simulations, the precipitation patterns do not match the observed VN distribution closely. In particular, high precipitation is observed around Hellas and in Arabia Terra (where Noachian-era VNs are scarce). However, low precipitation is observed in Margaritifer Sinus (where the integrated VN drainage density is nonetheless high). 

In the cold and icy simulations, spatial variations in supply rates are large, but the match between ice/snow accumulation and the VN distribution is closer. This is especially true at the most probable obliquity value predicted for the late Noachian, {where statistically significant spatial correlation is found between the two datasets.} At obliquities less than around $20^\circ$ and pressures less than around 0.5~bar, both \ce{CO2} and \ce{H2O} collapse at the poles, and the surface likely becomes too dry at equatorial latitudes for any significant melting or runoff to occur. Formation of a large south polar \ce{H2O} cap at low obliquity but higher pressure was already studied in \cite{Wordsworth2013a}, and may be linked to the late Noachian -- early Hesperian-era Dorsa Argentea Formation \citep{HeadPratt2001,Kress2015}.

Our results have implications for NASA's Curiosity mission. Specifically, we find relatively high snow accumulation rates in the Aeolis quadrangle in the cold, icy case, demonstrating that an ice-free northern ocean is not required to supply \ce{H2O} to Gale crater. A mainly cold scenario for fluvial alteration in Gale crater is consistent with \emph{in-situ} geochemical analyses of sedimentary material in the Yellowknife Bay formation \citep{McLennan14,Grotzinger14}.

Achieving the necessary transient warming to cause the required erosion through physically plausible mechanisms is an ongoing challenge for the cold scenario, although we have demonstrated that some melting is possible due to combinations of volcanism, dust radiative forcing and orbital variations. Impacts represent another major potential melting mechanism that by nature are also transient (e.g., \cite{Toon2010}). A warm, wet early Mars requires an anomalously luminous young Sun or intense greenhouse warming from an unknown source, and hence is less plausible from a pure climate physics perspective.

Clearly, there are many ways in which this work could be extended. The parametrizations we use for cloud microphysics and precipitation were chosen for their simplicity and independence from Earth-tuned parameters. It would be useful in future to compare them with more sophisticated schemes such as are currently used for Earth climate studies (e.g., \cite{Khairoutdinov2001}). In addition, it has been suggested recently that the two-stream approach to radiative transfer calculations (which we use) may be inaccurate for \ce{CO2} cloud scattering \citep{Kitzmann2013}. If this is the case, it would mean that our peak temperatures in Fig.~\ref{fig:melting} are slightly overestimated. It would of course also mean that warm, wet conditions in a \ce{CO2}-dominated atmosphere are even harder to achieve. This issue is not critical to our results, but it should be settled in the future using a 3D multiple-stream radiative transfer code.

For the cold scenario, future modeling work must focus on the construction of self-consistent scenarios for  melting episodes to explain the required erosion quantitatively. {More sophisticated modeling of hydrology on a cold early Mars (e.g., \cite{Fastook2015}) will also lead to a better understanding of the long-term evolution of water on the surface and beneath it.} Further work should also be performed to reconcile the late Noachian-era equatorial geomorphology with the southern Hesperian-era Dorsa Argentea Formation in future.

Finally, because our simulations make specific predictions for the spatial distribution of VNs and other fluvial features under different early climate scenarios, they strongly motivate further geological investigations. In the future, we suggest that special attention should be paid to morphological and geochemical analyses of Noachian terrain where predictions differ the most between warm, wet and cold, icy scenarios (e.g., in Arabia Terra and the Margaritifer Sinus quadrangle). More detailed orbital or in-situ analysis of these regions will lead to further insight into early Mars' global climate, and hence ultimately the question of whether there were ever long-term  conditions that could have allowed a surface biosphere to flourish.

\section{Acknowledgements}

The authors thank B. Ehlmann and a second anonymous reviewer for constructive comments that significantly improved the quality of the manuscript. We also thank Ehouarn Millour for help in setting up the parallelized version of the generic model that was used for some of the simulations. The computations in this paper were run on the Odyssey cluster supported by the FAS Division of Science, Research Computing Group at Harvard University. All data and code used to produce the figures in this article are available from the lead author on request (contact email rwordsworth@seas.harvard.edu). 


\begin{thebibliography}{79}
\providecommand{\natexlab}[1]{#1}
\providecommand{\url}[1]{\texttt{#1}}
\expandafter\ifx\csname urlstyle\endcsname\relax
  \providecommand{\doi}[1]{doi: #1}\else
  \providecommand{\doi}{doi: \begingroup \urlstyle{rm}\Url}\fi

\bibitem[{Baranov} et~al.(2004){Baranov}, {Lafferty}, and
  {Fraser}]{Baranov2004}
Y.~I. {Baranov}, W.~J. {Lafferty}, and G.~T. {Fraser}.
\newblock {Infrared spectrum of the continuum and dimer absorption in the
  vicinity of the O\_{2} vibrational fundamental in O\_{2}/CO\_{2} mixtures}.
\newblock \emph{Journal of Molecular Spectroscopy}, 228:\penalty0 432--440,
  December 2004.
\newblock \doi{10.1016/j.jms.2004.04.010}.

\bibitem[{Bibring} et~al.(2006){Bibring}, {Langevin}, {Mustard}, {Poulet},
  {Arvidson}, {Gendrin}, {Gondet}, {Mangold}, {Pinet}, and
  {Forget}]{Bibring2006}
{J.-P.} {Bibring}, Y.~{Langevin}, J.~F. {Mustard}, F.~{Poulet}, R.~{Arvidson},
  A.~{Gendrin}, B.~{Gondet}, N.~{Mangold}, P.~{Pinet}, and F.~{Forget}.
\newblock {Global Mineralogical and Aqueous Mars History Derived from
  OMEGA/Mars Express Data}.
\newblock \emph{Science}, 312:\penalty0 400--404, April 2006.
\newblock \doi{10.1126/science.1122659}.

\bibitem[Boucher et~al.(1995)Boucher, Le~Treut, and Baker]{Boucher1995}
O.~Boucher, H.~Le~Treut, and M.~B. Baker.
\newblock Precipitation and radiation modeling in a general circulation model:
  Introduction of cloud microphysical processes.
\newblock \emph{Journal of Geophysical Research: Atmospheres (1984--2012)},
  100\penalty0 (D8):\penalty0 16395--16414, 1995.

\bibitem[{Carr}(1996)]{Carr:96}
M.~H. {Carr}.
\newblock \emph{{Water on Mars}}.
\newblock New York: Oxford University Press, |c1996, 1996.

\bibitem[Carr and Head(2014)]{Carr2014}
M.~H. Carr and J.~W. Head.
\newblock Martian unbound water inventories: Changes with time.
\newblock In \emph{Lunar and Planetary Institute Science Conference Abstracts},
  volume~45, page 1427, 2014.

\bibitem[{Clifford} and {Parker}(2001)]{Clifford2001}
S.~M. {Clifford} and T.~J. {Parker}.
\newblock {The evolution of the martian hydrosphere: Implications for the fate
  of a primordial ocean and the current state of the northern plains}.
\newblock \emph{Icarus}, 154:\penalty0 40--79, November 2001.
\newblock \doi{10.1006/icar.2001.6671}.

\bibitem[Clough et~al.(1989)Clough, Kneizys, and Davies]{Clough1989}
S.~A. Clough, F.~X. Kneizys, and R.~W. Davies.
\newblock Line shape and the water vapor continuum.
\newblock \emph{Atmospheric Research}, 23\penalty0 (3-4):\penalty0 229 -- 241,
  1989.
\newblock ISSN 0169-8095.

\bibitem[Conway et~al.(1996)Conway, Gades, and Raymond]{Conway1996}
H.~Conway, A.~Gades, and C.~F. Raymond.
\newblock Albedo of dirty snow during conditions of melt.
\newblock \emph{Water Resources Research}, 32\penalty0 (6):\penalty0
  1713--1718, 1996.

\bibitem[Craddock and Howard(2002)]{Craddock2002}
Robert~A Craddock and Alan~D Howard.
\newblock The case for rainfall on a warm, wet early mars.
\newblock \emph{Journal of Geophysical Research}, 107\penalty0 (E11):\penalty0
  5111, 2002.

\bibitem[{di Achille} and {Hynek}(2010)]{diAchille2010}
G.~{di Achille} and B.~M. {Hynek}.
\newblock {Ancient ocean on Mars supported by global distribution of deltas and
  valleys}.
\newblock \emph{Nature Geoscience}, 3:\penalty0 459--463, July 2010.
\newblock \doi{10.1038/ngeo891}.

\bibitem[{Ehlmann} et~al.(2011){Ehlmann}, {Mustard}, {Murchie}, {Bibring},
  {Meunier}, {Fraeman}, and {Langevin}]{Ehlmann2011}
B.~L. {Ehlmann}, J.~F. {Mustard}, S.~L. {Murchie}, J.-P. {Bibring},
  A.~{Meunier}, A.~A. {Fraeman}, and Y.~{Langevin}.
\newblock {Subsurface water and clay mineral formation during the early history
  of Mars}.
\newblock \emph{Nature}, 479:\penalty0 53--60, November 2011.
\newblock \doi{10.1038/nature10582}.

\bibitem[Fair{\'e}n et~al.(2009)Fair{\'e}n, Davila, Gago-Duport, Amils, and
  McKay]{Fairen2009}
Alberto~G Fair{\'e}n, Alfonso~F Davila, Luis Gago-Duport, Ricardo Amils, and
  Christopher~P McKay.
\newblock Stability against freezing of aqueous solutions on early mars.
\newblock \emph{Nature}, 459\penalty0 (7245):\penalty0 401--404, 2009.

\bibitem[{Fassett} and {Head}(2008{\natexlab{a}})]{Fassett2008}
C.~I. {Fassett} and J.~W. {Head}.
\newblock {The timing of martian valley network activity: Constraints from
  buffered crater counting}.
\newblock \emph{Icarus}, 195:\penalty0 61--89, May 2008{\natexlab{a}}.
\newblock \doi{10.1016/j.icarus.2007.12.009}.

\bibitem[{Fassett} and {Head}(2008{\natexlab{b}})]{Fassett2008a}
C.~I. {Fassett} and J.~W. {Head}.
\newblock {Valley network-fed, open-basin lakes on Mars: Distribution and
  implications for Noachian surface and subsurface hydrology}.
\newblock \emph{Icarus}, 198:\penalty0 37--56, November 2008{\natexlab{b}}.
\newblock \doi{10.1016/j.icarus.2008.06.016}.

\bibitem[{Fassett} and {Head}(2011)]{Fassett2011}
C.~I. {Fassett} and J.~W. {Head}.
\newblock {Sequence and timing of conditions on early Mars}.
\newblock \emph{Icarus}, 211:\penalty0 1204--1214, February 2011.
\newblock \doi{10.1016/j.icarus.2010.11.014}.

\bibitem[Fastook and Head(2014)]{Fastook2015}
J.~L. Fastook and J.~W. Head.
\newblock Glaciation in the late noachian icy highlands: Ice accumulation,
  distribution, flow rates, basal melting, and top-down melting rates and
  patterns.
\newblock \emph{Planetary and Space Science}, 2014.

\bibitem[{Forget} and {Pierrehumbert}(1997)]{Forget1997}
F.~{Forget} and R.~T. {Pierrehumbert}.
\newblock {Warming Early Mars with Carbon Dioxide Clouds That Scatter Infrared
  Radiation}.
\newblock \emph{Science}, 278:\penalty0 1273--+, November 1997.

\bibitem[Forget et~al.(2013)Forget, Wordsworth, Millour, Madeleine, Kerber,
  Leconte, Marcq, and Haberle]{Forget2013}
F.~Forget, R.~D. Wordsworth, E.~Millour, J.-B. Madeleine, L.~Kerber,
  J.~Leconte, E.~Marcq, and R.~M. Haberle.
\newblock 3d modelling of the early martian climate under a denser {CO2}
  atmosphere: Temperatures and {CO2} ice clouds.
\newblock \emph{Icarus}, 2013.

\bibitem[Gill(1980)]{Gill1980}
A.~E. Gill.
\newblock Some simple solutions for heat-induced tropical circulation.
\newblock \emph{Quarterly Journal of the Royal Meteorological Society},
  106\penalty0 (449):\penalty0 447--462, 1980.

\bibitem[{Gough}(1981)]{Gough1981}
D.~O. {Gough}.
\newblock {Solar interior structure and luminosity variations}.
\newblock \emph{Solar Physics}, 74:\penalty0 21--34, November 1981.

\bibitem[Greeley and Guest(1987)]{Greeley1987}
R.~Greeley and J.~Guest.
\newblock \emph{Geologic map of the eastern equatorial region of Mars}.
\newblock Geological Survey (US), 1987.

\bibitem[{Grott} et~al.(2011){Grott}, {Morschhauser}, {Breuer}, and
  {Hauber}]{Grott2011}
M.~{Grott}, A.~{Morschhauser}, D.~{Breuer}, and E.~{Hauber}.
\newblock {Volcanic outgassing of CO$_{2}$ and H$_{2}$O on Mars}.
\newblock \emph{Earth and Planetary Science Letters}, 308:\penalty0 391--400,
  August 2011.
\newblock \doi{10.1016/j.epsl.2011.06.014}.

\bibitem[Grotzinger et~al.(2014)Grotzinger, Sumner, Kah, Stack, Gupta, Edgar,
  Rubin, Lewis, Schieber, Mangold, Milliken, Conrad, DesMarais, Farmer,
  Siebach, Calef, Hurowitz, McLennan, Ming, Vaniman, Crisp, Vasavada, Edgett,
  Malin, Blake, Gellert, Mahaffy, Wiens, Maurice, Grant, Wilson, Anderson,
  Beegle, Arvidson, Hallet, Sletten, Rice, Bell, Griffes, Ehlmann, Anderson,
  Bristow, Dietrich, Dromart, Eigenbrode, Fraeman, Hardgrove, Herkenhoff,
  Jandura, Kocurek, Lee, Leshin, Leveille, Limonadi, Maki, McCloskey, Meyer,
  Minitti, Newsom, Oehler, Okon, Palucis, Parker, Rowland, Schmidt, Squyres,
  Steele, Stolper, Summons, Treiman, Williams, Yingst, and Team]{Grotzinger14}
J.~P. Grotzinger, D.~Y. Sumner, L.~C. Kah, K.~Stack, S.~Gupta, L.~Edgar,
  D.~Rubin, K.~Lewis, J.~Schieber, N.~Mangold, R.~Milliken, P.~G. Conrad,
  D.~DesMarais, J.~Farmer, K.~Siebach, F.~Calef, J.~Hurowitz, S.~M. McLennan,
  D.~Ming, D.~Vaniman, J.~Crisp, A.~Vasavada, K.~S. Edgett, M.~Malin, D.~Blake,
  R.~Gellert, P.~Mahaffy, R.~C. Wiens, S.~Maurice, J.~A. Grant, S.~Wilson,
  R.~C. Anderson, L.~Beegle, R.~Arvidson, B.~Hallet, R.~S. Sletten, M.~Rice,
  J.~Bell, J.~Griffes, B.~Ehlmann, R.~B. Anderson, T.~F. Bristow, W.~E.
  Dietrich, G.~Dromart, J.~Eigenbrode, A.~Fraeman, C.~Hardgrove, K.~Herkenhoff,
  L.~Jandura, G.~Kocurek, S.~Lee, L.~A. Leshin, R.~Leveille, D.~Limonadi,
  J.~Maki, S.~McCloskey, M.~Meyer, M.~Minitti, H.~Newsom, D.~Oehler, A.~Okon,
  M.~Palucis, T.~Parker, S.~Rowland, M.~Schmidt, S.~Squyres, A.~Steele,
  E.~Stolper, R.~Summons, A.~Treiman, R.~Williams, A.~Yingst, and MSL~Science
  Team.
\newblock A habitable fluvio-lacustrine environment at yellowknife bay, gale
  crater, mars.
\newblock \emph{Science}, 343\penalty0 (6169), 2014.
\newblock \doi{10.1126/science.1242777}.
\newblock URL
  \url{http://www.sciencemag.org/content/343/6169/1242777.abstract}.

\bibitem[{Gruszka} and {Borysow}(1998)]{Gruszka1998}
M.~{Gruszka} and A.~{Borysow}.
\newblock {Computer simulation of the far infrared collision induced absorption
  spectra of gaseous CO2}.
\newblock \emph{Molecular Physics}, 93:\penalty0 1007--1016, 1998.
\newblock \doi{10.1080/002689798168709}.

\bibitem[Halevy and {Head}(2014)]{Halevy2014}
I.~Halevy and J.~W. {Head}.
\newblock Episodic warming of early mars by punctuated volcanism.
\newblock \emph{Nature Geoscience}, 2014.

\bibitem[{Head} and {Pratt}(2001)]{HeadPratt2001}
J.~W. {Head} and S.~{Pratt}.
\newblock {Extensive Hesperian-aged south polar ice sheet on Mars: Evidence for
  massive melting and retreat, and lateral flow and ponding of meltwater}.
\newblock \emph{Journal of Geophysical Research}, 106:\penalty0 12275--12300,
  June 2001.
\newblock \doi{10.1029/2000JE001359}.

\bibitem[{Head} et~al.(2002){Head}, {Kreslavsky}, and {Pratt}]{Head2002}
J.~W. {Head}, M.~A. {Kreslavsky}, and S.~{Pratt}.
\newblock {Northern lowlands of Mars: Evidence for widespread volcanic flooding
  and tectonic deformation in the Hesperian Period}.
\newblock \emph{Journal of Geophysical Research (Planets)}, 107:\penalty0 5003,
  January 2002.
\newblock \doi{10.1029/2000JE001445}.

\bibitem[Head and Marchant(2014)]{Head2014}
James~W Head and David~R Marchant.
\newblock The climate history of early mars: insights from the antarctic
  mcmurdo dry valleys hydrologic system.
\newblock \emph{Antarctic Science}, 26\penalty0 (06):\penalty0 774--800, 2014.

\bibitem[Hirschmann and Withers(2008)]{HirschmannWithers2008}
M.~M. Hirschmann and A.~C. Withers.
\newblock Ventilation of {CO2} from a reduced mantle and consequences for the
  early martian greenhouse.
\newblock \emph{Earth and Planetary Science Letters}, 270\penalty0
  (1):\penalty0 147--155, 2008.

\bibitem[{Hoke} et~al.(2011){Hoke}, {Hynek}, and Tucker]{Hoke2011}
M.~R.~T. {Hoke}, B.~M. {Hynek}, and G.~E. Tucker.
\newblock Formation timescales of large martian valley networks.
\newblock \emph{Earth and Planetary Science Letters}, 312\penalty0
  (1):\penalty0 1--12, 2011.

\bibitem[Holton and Hakim(2013)]{Holton2013}
J.~R. Holton and G.~J. Hakim.
\newblock \emph{An introduction to dynamic meteorology}.
\newblock Academic press, 2013.

\bibitem[Hudson and Yum(2002)]{Hudson2002}
James~G Hudson and Seong~Soo Yum.
\newblock Cloud condensation nuclei spectra and polluted and clean clouds over
  the indian ocean.
\newblock \emph{Journal of Geophysical Research: Atmospheres (1984--2012)},
  107\penalty0 (D19):\penalty0 INX2--21, 2002.

\bibitem[{Hynek} et~al.(2010){Hynek}, {Beach}, and {Hoke}]{Hynek2010}
B.~M. {Hynek}, M.~{Beach}, and M.~R.~T. {Hoke}.
\newblock {Updated global map of Martian valley networks and implications for
  climate and hydrologic processes}.
\newblock \emph{Journal of Geophysical Research (Planets)}, 115:\penalty0
  E09008, September 2010.
\newblock \doi{10.1029/2009JE003548}.

\bibitem[Kadish et~al.(2009)Kadish, Barlow, and Head]{Kadish2009}
S.~J. Kadish, N.~G. Barlow, and J.~W. Head.
\newblock Latitude dependence of martian pedestal craters: Evidence for a
  sublimation-driven formation mechanism.
\newblock \emph{Journal of Geophysical Research: Planets (1991--2012)},
  114\penalty0 (E10), 2009.

\bibitem[{Kasting}(1991)]{Kasting1991}
J.~F. {Kasting}.
\newblock {CO2 condensation and the climate of early Mars}.
\newblock \emph{Icarus}, 94:\penalty0 1--13, November 1991.

\bibitem[{Kerber} et~al.(2015){Kerber}, {Forget}, and {Wordsworth}]{Kerber2015}
L.~{Kerber}, F.~{Forget}, and R.~D. {Wordsworth}.
\newblock Sulfur in the early martian atmosphere revisited: Experiments with a
  3-d global climate model.
\newblock \emph{Icarus}, 2015.

\bibitem[Khairoutdinov and Randall(2001)]{Khairoutdinov2001}
Marat~F Khairoutdinov and David~A Randall.
\newblock A cloud resolving model as a cloud parameterization in the ncar
  community climate system model: Preliminary results.
\newblock \emph{Geophysical Research Letters}, 28\penalty0 (18):\penalty0
  3617--3620, 2001.

\bibitem[Kite et~al.(2014)Kite, Williams, Lucas, and Aharonson]{Kite2014}
E.~S. Kite, J.-P. Williams, A.~Lucas, and O.~Aharonson.
\newblock Low palaeopressure of the martian atmosphere estimated from the size
  distribution of ancient craters.
\newblock \emph{Nature Geoscience}, 7\penalty0 (5):\penalty0 335--339, 2014.

\bibitem[{Kitzmann} et~al.(2013){Kitzmann}, {Patzer}, and
  {Rauer}]{Kitzmann2013}
D.~{Kitzmann}, A.~B.~C. {Patzer}, and H.~{Rauer}.
\newblock {Clouds in the atmospheres of extrasolar planets. IV. On the
  scattering greenhouse effect of CO$_{2}$ ice particles: Numerical radiative
  transfer studies}.
\newblock \emph{Astronomy \& Astrophysics}, 557:\penalty0 A6, September 2013.
\newblock \doi{10.1051/0004-6361/201220025}.

\bibitem[Kress and Head(2015)]{Kress2015}
Ailish~M. Kress and James~W. Head.
\newblock Late noachian and early hesperian ridge systems in the south
  circumpolar dorsa argentea formation, mars: Evidence for two stages of
  melting of an extensive late noachian ice sheet.
\newblock \emph{Planetary and Space Science}, 109--110\penalty0 (0):\penalty0 1
  -- 20, 2015.
\newblock ISSN 0032-0633.
\newblock \doi{http://dx.doi.org/10.1016/j.pss.2014.11.025}.

\bibitem[{Laskar} et~al.(2004){Laskar}, {Correia}, {Gastineau}, {Joutel},
  {Levrard}, and {Robutel}]{Laskar2004}
J.~{Laskar}, A.~C.~M. {Correia}, M.~{Gastineau}, F.~{Joutel}, B.~{Levrard}, and
  P.~{Robutel}.
\newblock {Long term evolution and chaotic diffusion of the insolation
  quantities of Mars}.
\newblock \emph{Icarus}, 170:\penalty0 343--364, August 2004.
\newblock \doi{10.1016/j.icarus.2004.04.005}.

\bibitem[Le~Treut and Li(1991)]{LeTreut1991}
Herv\'e Le~Treut and Zhao-Xin Li.
\newblock Sensitivity of an atmospheric general circulation model to prescribed
  sst changes: feedback effects associated with the simulation of cloud optical
  properties.
\newblock \emph{Climate Dynamics}, 5:\penalty0 175--187, 1991.
\newblock ISSN 0930-7575.
\newblock URL \url{http://dx.doi.org/10.1007/BF00251808}.
\newblock 10.1007/BF00251808.

\bibitem[Lindzen and Hou(1988)]{Lindzen1988}
R.~S. Lindzen and A.~V. Hou.
\newblock Hadley circulations for zonally averaged heating centered off the
  equator.
\newblock \emph{Journal of the Atmospheric Sciences}, 45\penalty0
  (17):\penalty0 2416--2427, 1988.

\bibitem[{Madeleine} et~al.(2009){Madeleine}, {Forget}, {Head}, {Levrard},
  {Montmessin}, and {Millour}]{Madeleine2009}
J.-B. {Madeleine}, F.~{Forget}, J.~W. {Head}, B.~{Levrard}, F.~{Montmessin},
  and E.~{Millour}.
\newblock {Amazonian northern mid-latitude glaciation on Mars: A proposed
  climate scenario}.
\newblock \emph{Icarus}, 203:\penalty0 390--405, October 2009.
\newblock \doi{10.1016/j.icarus.2009.04.037}.

\bibitem[Manabe(1969)]{Manabe1969}
Syukuro Manabe.
\newblock Climate and the ocean circulation 1: The atmospheric circulation and
  the hydrology of the earth's surface.
\newblock \emph{Monthly Weather Review}, 97\penalty0 (11):\penalty0 739--774,
  1969.

\bibitem[Mangold et~al.(2004)Mangold, Quantin, Ansan, Delacourt, and
  Allemand]{Mangold2004}
N.~Mangold, C.~Quantin, V.~Ansan, C.~Delacourt, and P.~Allemand.
\newblock Evidence for precipitation on mars from dendritic valleys in the
  valles marineris area.
\newblock \emph{Science}, 305\penalty0 (5680):\penalty0 78--81, 2004.

\bibitem[McLennan et~al.(2014)McLennan, Anderson, Bell, Bridges, Calef,
  Campbell, Clark, Clegg, Conrad, Cousin, Des~Marais, Dromart, Dyar, Edgar,
  Ehlmann, Fabre, Forni, Gasnault, Gellert, Gordon, Grant, Grotzinger, Gupta,
  Herkenhoff, Hurowitz, King, Le~Mou{\'e}lic, Leshin, L{\'e}veill{\'e}, Lewis,
  Mangold, Maurice, Ming, Morris, Nachon, Newsom, Ollila, Perrett, Rice,
  Schmidt, Schwenzer, Stack, Stolper, Sumner, Treiman, VanBommel, Vaniman,
  Vasavada, Wiens, Yingst, and Team]{McLennan14}
S.~M. McLennan, R.~B. Anderson, J.~F. Bell, J.~C. Bridges, F.~Calef, J.~L.
  Campbell, B.~C. Clark, S.~Clegg, P.~Conrad, A.~Cousin, D.~J. Des~Marais,
  G.~Dromart, M.~D. Dyar, L.~A. Edgar, B.~L. Ehlmann, C.~Fabre, O.~Forni,
  O.~Gasnault, R.~Gellert, S.~Gordon, J.~A. Grant, J.~P. Grotzinger, S.~Gupta,
  K.~E. Herkenhoff, J.~A. Hurowitz, P.~L. King, S.~Le~Mou{\'e}lic, L.~A.
  Leshin, R.~L{\'e}veill{\'e}, K.~W. Lewis, N.~Mangold, S.~Maurice, D.~W. Ming,
  R.~V. Morris, M.~Nachon, H.~E. Newsom, A.~M. Ollila, G.~M. Perrett, M.~S.
  Rice, M.~E. Schmidt, S.~P. Schwenzer, K.~Stack, E.~M. Stolper, D.~Y. Sumner,
  A.~H. Treiman, S.~VanBommel, D.~T. Vaniman, A.~Vasavada, R.~C. Wiens, R.~A.
  Yingst, and MSL~Science Team.
\newblock Elemental geochemistry of sedimentary rocks at yellowknife bay, gale
  crater, mars.
\newblock \emph{Science}, 343\penalty0 (6169), 2014.
\newblock \doi{10.1126/science.1244734}.
\newblock URL
  \url{http://www.sciencemag.org/content/343/6169/1244734.abstract}.

\bibitem[Minton and Malhotra(2007)]{MintonMalhotra2007}
David~A Minton and Renu Malhotra.
\newblock Assessing the massive young sun hypothesis to solve the warm young
  earth puzzle.
\newblock \emph{The Astrophysical Journal}, 660\penalty0 (2):\penalty0 1700,
  2007.

\bibitem[Mischna et~al.(2013)Mischna, Baker, Milliken, Richardson, and
  Lee]{Mischna2013}
M.~A. Mischna, V.~Baker, R.~Milliken, M.~Richardson, and C.~Lee.
\newblock Effects of obliquity and water vapor/trace gas greenhouses in the
  early martian climate.
\newblock \emph{Journal of Geophysical Research: Planets}, 2013.

\bibitem[Mitchell et~al.(2006)Mitchell, Pierrehumbert, Frierson, and
  Caballero]{Mitchell2006}
Jonathan~L Mitchell, Raymond~T Pierrehumbert, Dargan~MW Frierson, and Rodrigo
  Caballero.
\newblock The dynamics behind titan's methane clouds.
\newblock \emph{Proceedings of the National Academy of Sciences}, 103\penalty0
  (49):\penalty0 18421--18426, 2006.

\bibitem[{Mustard} et~al.(2008){Mustard}, {Murchie}, {Pelkey}, {Ehlmann},
  {Milliken}, {Grant}, {Bibring}, {Poulet}, {Bishop}, {Dobrea}, {Roach},
  {Seelos}, {Arvidson}, {Wiseman}, {Green}, {Hash}, {Humm}, {Malaret},
  {McGovern}, {Seelos}, {Clancy}, {Clark}, {Marais}, {Izenberg}, {Knudson},
  {Langevin}, {Martin}, {McGuire}, {Morris}, {Robinson}, {Roush}, {Smith},
  {Swayze}, {Taylor}, {Titus}, and {Wolff}]{Mustard2008}
J.~F. {Mustard}, S.~L. {Murchie}, S.~M. {Pelkey}, B.~L. {Ehlmann}, R.~E.
  {Milliken}, J.~A. {Grant}, J.-P. {Bibring}, F.~{Poulet}, J.~{Bishop}, E.~N.
  {Dobrea}, L.~{Roach}, F.~{Seelos}, R.~E. {Arvidson}, S.~{Wiseman},
  R.~{Green}, C.~{Hash}, D.~{Humm}, E.~{Malaret}, J.~A. {McGovern},
  K.~{Seelos}, T.~{Clancy}, R.~{Clark}, D.~D. {Marais}, N.~{Izenberg},
  A.~{Knudson}, Y.~{Langevin}, T.~{Martin}, P.~{McGuire}, R.~{Morris},
  M.~{Robinson}, T.~{Roush}, M.~{Smith}, G.~{Swayze}, H.~{Taylor}, T.~{Titus},
  and M.~{Wolff}.
\newblock {Hydrated silicate minerals on Mars observed by the Mars
  Reconnaissance Orbiter CRISM instrument}.
\newblock \emph{Nature}, 454:\penalty0 305--309, July 2008.
\newblock \doi{10.1038/nature07097}.

\bibitem[Nakajima et~al.(1992)Nakajima, Hayashi, and Abe]{Nakajima1992}
Shinichi Nakajima, Yoshi-Yuki Hayashi, and Yukata Abe.
\newblock A study on the {\"e}runaway greenhouse effect{\'\i}with a
  one-dimensional radiative--convective equilibrium model.
\newblock \emph{J. Atmos. Sci}, 49:\penalty0 2256--2266, 1992.

\bibitem[{Phillips} et~al.(2001){Phillips}, {Zuber}, {Solomon}, {Golombek},
  {Jakosky}, {Banerdt}, {Smith}, {Williams}, {Hynek}, {Aharonson}, and
  {Hauck}]{Phillips2001}
R.~J. {Phillips}, M.~T. {Zuber}, S.~C. {Solomon}, M.~P. {Golombek}, B.~M.
  {Jakosky}, W.~B. {Banerdt}, D.~E. {Smith}, R.~M.~E. {Williams}, B.~M.
  {Hynek}, O.~{Aharonson}, and S.~A. {Hauck}.
\newblock {Ancient geodynamics and global-scale hydrology on Mars}.
\newblock \emph{Science}, 291:\penalty0 2587--2591, March 2001.
\newblock \doi{10.1126/science.1058701}.

\bibitem[Pierrehumbert(2011)]{Pierrehumbert2011BOOK}
R.~T. Pierrehumbert.
\newblock \emph{Principles of Planetary Climate}.
\newblock Cambridge University Press, 2011.
\newblock ISBN 9780521865562.
\newblock URL \url{http://books.google.com/books?id=bO\_U8f5pVR8C}.

\bibitem[Pierrehumbert and Erlick(1998)]{Pierrehumbert1998}
R.~T. Pierrehumbert and C.~Erlick.
\newblock On the scattering greenhouse effect of co2 ice clouds.
\newblock \emph{Journal of the atmospheric sciences}, 55\penalty0
  (10):\penalty0 1897--1903, 1998.

\bibitem[{Pollack} et~al.(1987){Pollack}, {Kasting}, {Richardson}, and
  {Poliakoff}]{Pollack1987}
J.~B. {Pollack}, J.~F. {Kasting}, S.~M. {Richardson}, and K.~{Poliakoff}.
\newblock {The case for a wet, warm climate on early Mars}.
\newblock \emph{Icarus}, 71:\penalty0 203--224, August 1987.
\newblock \doi{10.1016/0019-1035(87)90147-3}.

\bibitem[{Postawko} and {Kuhn}(1986)]{Postawko1986}
S.~E. {Postawko} and W.~R. {Kuhn}.
\newblock {Effect of the greenhouse gases (CO$_{2}$, H$_{2}$O, SO$_{2}$) on
  martian paleoclimate}.
\newblock \emph{Journal of Geophysical Research}, 91:\penalty0 431--D438,
  September 1986.

\bibitem[Ramirez et~al.(2014)Ramirez, Kopparapu, Zugger, Robinson, Freedman,
  and Kasting]{Ramirez2014}
R.~M. Ramirez, R.~Kopparapu, M.~E. Zugger, T.~D. Robinson, R.~Freedman, and
  J.~F. Kasting.
\newblock Warming early mars with {CO2} and {H2}.
\newblock \emph{Nature Geoscience}, 7\penalty0 (1):\penalty0 59--63, 2014.

\bibitem[Richardson and Mischna(2005)]{Richardson2005}
M.~I. Richardson and M.~A. Mischna.
\newblock Long-term evolution of transient liquid water on mars.
\newblock \emph{Journal of Geophysical Research: Planets (1991--2012)},
  110\penalty0 (E3), 2005.

\bibitem[{Rothman} et~al.(2009){Rothman}, {Gordon}, {Barbe}, {Benner},
  {Bernath}, {Birk}, {Boudon}, {Brown}, {Campargue}, {Champion}, {Chance},
  {Coudert}, {Dana}, {Devi}, {Fally}, {Flaud}, {Gamache}, {Goldman},
  {Jacquemart}, {Kleiner}, {Lacome}, {Lafferty}, {Mandin}, {Massie},
  {Mikhailenko}, {Miller}, {Moazzen-Ahmadi}, {Naumenko}, {Nikitin}, {Orphal},
  {Perevalov}, {Perrin}, {Predoi-Cross}, {Rinsland}, {Rotger}, {{\v S}ime{\v
  c}kov{\'a}}, {Smith}, {Sung}, {Tashkun}, {Tennyson}, {Toth}, {Vandaele}, and
  {Vander Auwera}]{Rothman2009}
L.~S. {Rothman}, I.~E. {Gordon}, A.~{Barbe}, D.~C. {Benner}, P.~F. {Bernath},
  M.~{Birk}, V.~{Boudon}, L.~R. {Brown}, A.~{Campargue}, J.-P. {Champion},
  K.~{Chance}, L.~H. {Coudert}, V.~{Dana}, V.~M. {Devi}, S.~{Fally}, J.-M.
  {Flaud}, R.~R. {Gamache}, A.~{Goldman}, D.~{Jacquemart}, I.~{Kleiner},
  N.~{Lacome}, W.~J. {Lafferty}, J.-Y. {Mandin}, S.~T. {Massie}, S.~N.
  {Mikhailenko}, C.~E. {Miller}, N.~{Moazzen-Ahmadi}, O.~V. {Naumenko}, A.~V.
  {Nikitin}, J.~{Orphal}, V.~I. {Perevalov}, A.~{Perrin}, A.~{Predoi-Cross},
  C.~P. {Rinsland}, M.~{Rotger}, M.~{{\v S}ime{\v c}kov{\'a}}, M.~A.~H.
  {Smith}, K.~{Sung}, S.~A. {Tashkun}, J.~{Tennyson}, R.~A. {Toth}, A.~C.
  {Vandaele}, and J.~{Vander Auwera}.
\newblock {The HITRAN 2008 molecular spectroscopic database}.
\newblock \emph{Journal of Quantitative Spectroscopy and Radiative Transfer},
  110:\penalty0 533--572, 2009.
\newblock \doi{10.1016/j.jqsrt.2009.02.013}.

\bibitem[{Sadourny}(1975)]{Sadourny1975}
R.~{Sadourny}.
\newblock {The Dynamics of Finite-Difference Models of the Shallow-Water
  Equations.}
\newblock \emph{Journal of Atmospheric Sciences}, 32:\penalty0 680--689, 1975.

\bibitem[{Sagan} and {Mullen}(1972)]{Saga:72}
C.~{Sagan} and G.~{Mullen}.
\newblock {Earth and Mars: Evolution of Atmospheres and Surface Temperatures}.
\newblock \emph{Science}, 177:\penalty0 52--56, 1972.
\newblock \doi{10.1126/science.177.4043.52}.

\bibitem[Scott and Tanaka(1986)]{Scott1986}
D.~H. Scott and K.~L. Tanaka.
\newblock \emph{Geologic map of the western equatorial region of Mars}.
\newblock Geological Survey (US), 1986.

\bibitem[{Segura} et~al.(2008){Segura}, {Toon}, and {Colaprete}]{Segura2008}
T.~L. {Segura}, O.~B. {Toon}, and A.~{Colaprete}.
\newblock {Modeling the environmental effects of moderate-sized impacts on
  Mars}.
\newblock \emph{Journal of Geophysical Research (Planets)}, 113:\penalty0
  E11007, November 2008.
\newblock \doi{10.1029/2008JE003147}.

\bibitem[Segura et~al.(2012)Segura, McKay, and Toon]{Segura2012}
Teresa~L Segura, Christopher~P McKay, and Owen~B Toon.
\newblock An impact-induced, stable, runaway climate on mars.
\newblock \emph{Icarus}, 220\penalty0 (1):\penalty0 144--148, 2012.

\bibitem[Soderblom et~al.(1973)Soderblom, Kreidler, and
  Masursky]{Soderblom1973}
L.~A. Soderblom, T.~J. Kreidler, and H.~Masursky.
\newblock Latitudinal distribution of a debris mantle on the martian surface.
\newblock \emph{Journal of Geophysical Research}, 78\penalty0 (20):\penalty0
  4117--4122, 1973.

\bibitem[{Squyres} and {Kasting}(1994)]{Squyres1994}
S.~W. {Squyres} and J.~F. {Kasting}.
\newblock {Early Mars: How Warm and How Wet?}
\newblock \emph{Science}, 265:\penalty0 744--749, August 1994.
\newblock \doi{10.1126/science.265.5173.744}.

\bibitem[Tanaka(1986)]{Tanaka1986}
K.~L. Tanaka.
\newblock The stratigraphy of mars.
\newblock \emph{Journal of Geophysical Research: Solid Earth (1978--2012)},
  91\penalty0 (B13):\penalty0 E139--E158, 1986.

\bibitem[Tanaka and Scott(1987)]{Tanaka1987}
K.~L. Tanaka and D.~H. Scott.
\newblock \emph{Geologic map of the polar regions of Mars}.
\newblock Geological Survey (US), 1987.

\bibitem[{Toon} et~al.(1989){Toon}, {McKay}, {Ackerman}, and
  {Santhanam}]{Toon1989}
O.~B. {Toon}, C.~P. {McKay}, T.~P. {Ackerman}, and K.~{Santhanam}.
\newblock {Rapid calculation of radiative heating rates and photodissociation
  rates in inhomogeneous multiple scattering atmospheres}.
\newblock \emph{Journal of Geophysical Research}, 94:\penalty0 16287--16301,
  November 1989.

\bibitem[{Toon} et~al.(2010){Toon}, {Segura}, and {Zahnle}]{Toon2010}
O.~B. {Toon}, T.~{Segura}, and K.~{Zahnle}.
\newblock {The formation of martian river valleys by impacts}.
\newblock \emph{Annual Review of Earth and Planetary Sciences}, 38:\penalty0
  303--322, May 2010.
\newblock \doi{10.1146/annurev-earth-040809-152354}.

\bibitem[Urata and Toon(2013)]{UrataToon2013}
R.~A. Urata and O.~B. Toon.
\newblock Simulations of the martian hydrologic cycle with a general
  circulation model: Implications for the ancient martian climate.
\newblock \emph{Icarus}, 2013.

\bibitem[Webster(1977)]{Webster1977}
P.~J. Webster.
\newblock The low-latitude circulation of mars.
\newblock \emph{Icarus}, 30\penalty0 (4):\penalty0 626--649, 1977.

\bibitem[Williams et~al.(2013)Williams, Grotzinger, Dietrich, Gupta, Sumner,
  Wiens, Mangold, Malin, Edgett, Maurice, et~al.]{Williams2013}
R.~M.~E. Williams, J.~P. Grotzinger, W.~E. Dietrich, S.~Gupta, D.~Y. Sumner,
  R.~C. Wiens, N.~Mangold, M.~C. Malin, K.~S. Edgett, S.~Maurice, et~al.
\newblock Martian fluvial conglomerates at gale crater.
\newblock \emph{Science}, 340\penalty0 (6136):\penalty0 1068--1072, 2013.

\bibitem[Wordsworth and Pierrehumbert(2013{\natexlab{a}})]{Wordsworth2013c}
R.~Wordsworth and R.~Pierrehumbert.
\newblock Hydrogen-nitrogen greenhouse warming in earth's early atmosphere.
\newblock \emph{Science}, 339\penalty0 (6115):\penalty0 64--67,
  2013{\natexlab{a}}.
\newblock \doi{10.1126/science.1225759}.
\newblock URL \url{http://www.sciencemag.org/content/339/6115/64.abstract}.

\bibitem[Wordsworth and Pierrehumbert(2013{\natexlab{b}})]{Wordsworth2013water}
R.~Wordsworth and R.~Pierrehumbert.
\newblock Water loss from terrestrial planets with {CO2}-rich atmospheres.
\newblock \emph{The Astrophysical Journal}, 778\penalty0 (2):\penalty0 154,
  2013{\natexlab{b}}.

\bibitem[{Wordsworth} et~al.(2010{\natexlab{a}}){Wordsworth}, {Forget}, and
  {Eymet}]{Wordsworth2010}
R.~{Wordsworth}, F.~{Forget}, and V.~{Eymet}.
\newblock {Infrared collision-induced and far-line absorption in dense CO2
  atmospheres}.
\newblock \emph{Icarus}, 210:\penalty0 992--997, December 2010{\natexlab{a}}.
\newblock \doi{10.1016/j.icarus.2010.06.010}.

\bibitem[{Wordsworth} et~al.(2010{\natexlab{b}}){Wordsworth}, {Forget},
  {Selsis}, {Madeleine}, {Millour}, and {Eymet}]{Wordsworth2010b}
R.~{Wordsworth}, F.~{Forget}, F.~{Selsis}, {J.-B.} {Madeleine}, E.~{Millour},
  and V.~{Eymet}.
\newblock {Is Gliese 581d habitable? Some constraints from radiative-convective
  climate modeling}.
\newblock \emph{Astronomy and Astrophysics}, 522:\penalty0 A22+,
  2010{\natexlab{b}}.
\newblock \doi{10.1051/0004-6361/201015053}.

\bibitem[Wordsworth et~al.(2013)Wordsworth, Forget, Millour, Head, Madeleine,
  and Charnay]{Wordsworth2013a}
R.~Wordsworth, F.~Forget, E.~Millour, J.~W. Head, J.-B. Madeleine, and
  B.~Charnay.
\newblock Global modelling of the early martian climate under a denser {CO2}
  atmosphere: Water cycle and ice evolution.
\newblock \emph{Icarus}, 222\penalty0 (1):\penalty0 1--19, 2013.

\end{thebibliography}

\end{article}

\begin{table}[h]
\centering
\caption{Parameters used in the warm and wet climate simulations. Bold indicates the values for the `standard' simulations.}
\begin{tabular}{cc}
\hline
\hline
Surface pressure  & \textbf{1}~bar  \\
Solar flux  & 441.1, \textbf{764.5}~W~m$^{-2}$   \\
Added atmospheric gray mass absorption coefficient & \textbf{0.0}, 2.0$\times10^{-4}$~m$^2$~kg$^{-1}$   \\
Orbital obliquity & 25.0$^\circ$, \textbf{41.8}$^\circ$    \\
Orbital eccentricity &   \textbf{0.0}     \\
Runoff threshold  & 1, \textbf{150}~kg~m$^{-2}$     \\
\hline
\hline
\end{tabular}\label{tab:params1}
\end{table}

\begin{table}[h]
\centering
\caption{Parameters used in the cold and icy climate simulations. Bold indicates the values for the `standard' simulations.}
\begin{tabular}{cc}
\hline
\hline
Surface pressure  & 0.125, \textbf{0.6}, 2~bar    \\
Solar flux  & \textbf{441.1}~W~m$^{-2}$, 470.5~W~m$^{-2}$,     \\
Orbital obliquity &   10.0$^\circ$, 25.0$^\circ$, \textbf{41.8$^\circ$}, 55.0$^\circ$    \\
Orbital eccentricity &   \textbf{0.0}, 0.125    \\
Atmospheric \ce{SO2}  &   \textbf{0.0}, 10.0~ppmv    \\
Dust optical depth at 0.7~$\mu$m &   \textbf{0.0}, 5.0    \\
\hline
\hline
\end{tabular}\label{tab:params2}
\end{table}

\end{document}